\newif\ifdraft
\draftfalse

\documentclass[]{amsart}

\usepackage{color,graphicx,array,tabularx,url,enumerate,rotating,fancyvrb,color}
\usepackage{cite}
\usepackage{fullpage}
\usepackage{algorithmicx}
\usepackage{algorithm}
\usepackage{caption}
\usepackage{subcaption}
\usepackage{algpseudocode}

\usepackage{cleveref}
 \AtBeginDocument{%
    \crefname{figure}{fig.}{figs.}%
    \Crefname{figure}{Fig.}{Figs.}%
 }

 \usepackage{pgfplots}
\usepackage{pgfplotstable}

\pgfplotsset{
    discard if/.style 2 args={
        x filter/.code={
            \edef\tempa{\thisrow{#1}}
            \edef\tempb{#2}
            \ifx\tempa\tempb
                
            \fi
        }
    },
    discard if not/.style 2 args={
        x filter/.code={
            \edef\tempa{\thisrow{#1}}
            \edef\tempb{#2}
            \ifx\tempa\tempb
            \else
                
            \fi
        }
    }
  }
  \graphicspath{{img/}}
\usepackage{epstopdf}			%%%%%%%%%%%%%%%%%%%%%%%%%%%%%%%%%%%%%%%%%%%%%%%%%%%%%%%%%%%%%%%%%%%%%%%%%%%%5
\usepackage[section]{placeins}

\newtheorem{theorem}{Theorem}[section]

\newtheorem{definition}[theorem]{Definition}

\ifdraft
\usepackage{framed}
\fi

\newcommand\snapflow{INRFlow}

\newcommand{\tor}{\textit{TOR}}

\newcommand{\knkthreecol}{blue!70!black}
\newcommand{\knkfourcol}{purple!25!white}
\newcommand{\ficonncol}{green!60!black}

\newcommand\set[1]{\{#1\}}

\newcommand\flr[2]{\left\lfloor \frac{#1}{#2} \right\rfloor}

\renewcommand\phi{\theta}

\renewcommand\ast{\star}

\title{The Stellar Transformation:\\ From Interconnection Networks to
  Datacenter Networks}

\author{Alejandro Erickson}
\author{Iain A. Stewart}

\address{School of Engineering and Computing Sciences\\
Durham University, Science Labs, South Road\\
Durham DH1 3LE, U.K.}

\author{Javier Navaridas}
\author{Abbas E. Kiasari}

\address{School of Computer Science\\
University of Manchester, Oxford Road\\
Manchester M13 9PL, U.K.}

\begin{document}

\maketitle

\begin{abstract}
The first dual-port server-centric datacenter network, FiConn, was introduced in 2009 and there are several others now in existence; however, the pool of topologies to choose from remains small.  We propose a new generic construction, the stellar transformation, that dramatically increases the size of this pool by facilitating the transformation of well-studied topologies from interconnection networks, along with their networking properties and routing algorithms, into viable dual-port server-centric datacenter network topologies.  We demonstrate that under our transformation, numerous interconnection networks yield datacenter network topologies with potentially good, and easily computable, baseline  properties. We instantiate our construction so as to apply it to generalized hypercubes and obtain the datacenter networks GQ$^\star$.  Our construction automatically yields routing algorithms for GQ$^\star$ and we empirically compare GQ$^\star$ (and its routing algorithms) with the established datacenter networks FiConn and DPillar (and their routing algorithms); this comparison is with respect to network throughput, latency, load balancing, fault-tolerance, and cost to build, and is with regard to all-to-all, many all-to-all, butterfly, and random traffic patterns.  We find that GQ$^\star$ outperforms both FiConn and DPillar (sometimes significantly so) and that there is substantial scope for our stellar transformation to yield new dual-port server-centric datacenter networks that are a considerable improvement on existing ones.
\end{abstract}

% \begin{keyword}
%   dual-port server-centric datacenter networks \sep %
%   fault tolerance  \sep %
%   generalized hypercubes \sep %
%   interconnection networks \sep %
%   performance evaluation \sep %
%   routing \sep %
%   topological properties %
% \end{keyword}

\section{Introduction}\label{sec:introduction}

The digital economy has taken the world by storm and completely
changed the way we interact, communicate, collaborate, and search for information.  The main driver of this change has been the rapid
penetration of cloud computing which has enabled a wide variety of
digital services, such as web search and online gaming, by offering
elastic, on-demand computing resources to digital service providers.
Indeed, the value of the global cloud computing market is estimated to be in excess of \$100 billion \cite{WuBuyya2012}.  Vital to this ecosystem of digital services is an underlying computing infrastructure based primarily in datacenters~\cite{ArmbrustFoxGriffith2010}.  With this sudden move to the cloud, the demand for increasingly large datacenters is growing rapidly~\cite{GuoWuTan2008}.

This demand has prompted a move away from traditional datacenter
designs, based on expensive high-density enterprise-level switches,
towards using commodity-off-the-shelf (COTS) hardware.  In their
production datacenters, major operators have primarily adopted (and
invented) ideas similar to Fat-Tree \cite{Al-FaresLoukissasVahdat2008}, Portland \cite{Niranjan-MysorePamborisFarrington2009}, and VL2 \cite{GreenbergHamiltonJain2009}; on the other hand, the research
community (several major operators included) maintains a diverse
economy of datacenter architectures and designs in order to meet
future demand
\cite{HamedazimiQaziGupta2014,LiuGaoWong2014a,GuoWuTan2008, FarringtonPorterRadhakrishnan2010,QuFangZhang2015,SinglaHongPopa2012}.
Indeed, the ``switch-centric'' datacenters currently used in
production datacenters have inherent scalability limitations and are
by no means a low-cost solution (see, \emph{e.g.}, 
\cite{GuoWuTan2008,LiuMuppalaVeeraraghavan2013, GuoChenLi2013,ChenGaoChen2016}).

One approach intended to help overcome these limitations is the
``server-centric'' architecture, the first examples of which are DCell
\cite{GuoWuTan2008} and BCube \cite{GuoLuLi2009}.  Whereas in a switch-centric datacenter network (DCN) there are no links joining pairs of servers, in a server-centric DCN there are no links joining pairs of switches. This server-centric restriction arises from the circumstance that the switches in a server-centric DCN act only as non-blocking ``dumb'' crossbars.  By offloading the
task of routing packets to the servers, the server-centric architecture leverages the
typically low utilisation of CPUs in datacenters to manage network
communication.  This can reduce both the number of switches used in a
DCN and the capabilities required of them.  In particular, the
switches route only locally, to their neighbouring servers, and
therefore have no need for large or fast routing tables.  Thus, a
server-centric DCN can potentially incorporate more servers and be both cheaper to
operate and to build (see \cite{PopaRatnasamyIannaccone2010} for a
more detailed discussion).  Furthermore, using servers (which are highly
programmable) rather than switches (which have
proprietary software and limited programmability) to route packets will potentially
accelerate research innovation \cite{LiGuoYang2014}. 

Since the advent of DCell and BCube, a range of server-centric DCNs have been proposed, some of which further restrict themselves to
requiring at most two ports per server, with FiConn \cite{LiGuoWu2011} and DPillar \cite{LiaoYinYin2012} being the most established of this genre. This dual-port restriction is motivated by the fact that many COTS
servers presently available for purchase, as well as servers in
existing datacenters, have two NIC ports (a primary and a backup port).
Dual-port server-centric DCNs are able to utilise such servers without
modification, thus making it possible to use some of the more basic
equipment (available for purchase or from existing datacenters) in a
server-centric DCN and thereby reduce the building costs. 

The server-centric DCN architecture provides a versatile design space,
as regards the network topology, evidenced perhaps by the sheer number
of fairly natural constructions proposed from 2008 to the present.  On
the other hand, this pool is small relative to the number of
interconnection networks found in the literature, \emph{i.e.}, highly
structured graphs with good networking properties.  One of the
challenges of identifying an interconnection network suitable for
conversion to a DCN topology, however, lies in the fact that the
literature on interconnection networks is focused primarily on graphs
whose nodes are homogeneous\footnote{We disregard the terminal nodes
  of indirect networks, which are not intrinsic to the topology.}, whereas in both a switch-centric and a server-centric DCN we have server-nodes and switch-nodes which have entirely different operational roles.
Some server-centric DCN topologies arise largely from outside the interconnection network
literature, \emph{e.g.}, DCell and FiConn, whilst others arise from transformations of
well-known interconnection networks, \emph{e.g.}, BCube and DPillar.

The transformations used to obtain BCube and DPillar take advantage of
certain sub-structures in the underlying base graphs of the interconnection networks in question (generalized hypercubes and wrapped butterfly networks, respectively) in order to create a
server-centric DCN that inherits beneficial networking
properties such as having a low diameter and fault-tolerant routing
algorithms.  The limitation, of course, is that not every
prospective base graph has the required sub-structures (cliques and
bicliques, respectively, in the cases of BCube and DPillar).  New methods of
transforming interconnection networks into server-centric DCNs may
therefore greatly enlarge the server-centric DCN design space by lowering the
structural requirements on potential base graphs.

It is with the construction of new dual-port server-centric DCNs that we are
concerned in this paper. In particular, we show how to systematically transform
interconnection networks, as base graphs, into dual-port server-centric DCNs,
which we refer to as \emph{stellar\/} DCNs. The stellar transformation is very
simple and widely applicable: the edges of the base graph are replaced with
3-paths, so that the nodes of the base graph become the switch-nodes of the
stellar DCN and the added nodes, interior to the 3-paths, become the
server-nodes (see \Cref{fig:knkstar}). By requiring very little of the base
graph in the way of structure, the stellar construction greatly increases the
pool of interconnection networks that can potentially serve as blueprints to
design dual-port server-centric DCN topologies.

We validate our construction in three ways: first, we prove
that various good networking properties of the base graph are
preserved under the stellar transformation;
second, we build a library of interconnection networks that
suit the stellar transformation; and third, we empirically evaluate
GQ$^\star$, an instantiation of a stellar DCN whose base graph
is a generalized hypercube, against both FiConn and DPillar, and we also compare GQ$^\star$ and its routing algorithm (inherited from generalized hypercubes) against what might be optimally possible in GQ$^\star$.

Our empirical results are extremely encouraging. We employ a comprehensive set of performance metrics so as to evaluate network
throughput, latency, load balancing capability, fault-tolerance, and cost to build, within the context of all-to-all, many all-to-all, butterfly, and random traffic patterns, and we show that GQ$^\star$ broadly outperforms both FiConn and DPillar as regards these metrics, sometimes significantly so. Highlights of these improvemenmts are as follows. In terms of aggregate bottleneck throughput (a primary metric as regards the evaluation of throughput in an all-to-all context), our DCN GQ$^\star$ improves upon both FiConn and DPillar (upon the former markedly so). As regards fault-tolerance, our DCN GQ$^\star$, with its fault-tolerant routing algorithm \emph{GQ$^\star$-routing\/} (inherited from generalized hypercubes), outperforms DPillar (and its fault-tolerant routing algorithm \emph{DPillarMP\/} from \cite{LiaoYinYin2012}) and competes with FiConn even when we simulate optimal fault-tolerant routing in FiConn (even though such a fault-tolerant routing algorithm has yet to be exhibited). Not only does \emph{GQ$^\star$-routing\/} (in GQ$^\star$) tolerate faults better than the respective routing algorithms in FiConn and DPillar, but when we make around 10\% of the links faulty and compare it with the optimal scenario in GQ$^\star$, \emph{GQ$^\star$-routing\/} provides around 95\% connectivity and generates paths that are, on average, only around 10\% longer than the shortest available paths. When we consider load balancing in GQ$^\star$, FiConn, and DPillar, with their respective routing algorithms \emph{GQ$^\star$-routing\/}, \emph{TOR\/}, and \emph{DPillarSP\/}  and under a variety of traffic patterns, we find that the situation in GQ$^\star$ is demonstrably improved over that in FiConn and DPillar (with DPillar performing particularly poorly), and that the improved load balancing in GQ$^\star$ in tandem with the generation of relatively short paths translates to potential latency savings.

However, we have only scratched the surface in terms of what might be possible as regards the translation of high-performance interconnection networks into dual-port server-centric DCNs in that we have applied our generic, stellar construction to only one family of interconnection networks so as to achieve encouraging results. In addition to our experiments, we demonstrate that there are numerous families of interconnection networks to which our construction might be applied. Whilst our results with generalized hypercubes are extremely positive, we feel that the generic nature of our construction has significant potential and scope for further application.

The rest of the paper is organized as follows.  In the next section,
we give an overview of the design space for dual-port server-centric
DCNs, along with related work, before defining our new generic construction
in Section~\ref{sec:generic} and proving that good networking properties of the underlying interconnection network translate to good networking properties of the stellar DCN.  We also instantiate our stellar construction in Section~\ref{sec:generic} so
as to generate the DCNs GQ$^\star$, and in Sections~\ref{sec:setup} and~\ref{sec:evaluation} we describe the methodology of our empirical evaluation and
the results of this investigation, respectively.  We close with some concluding remarks and
suggestions for future work in Section~\ref{sec:conclusion}. We refer the reader: to \cite{Diestel2012} for all standard graph-theoretic concepts; to \cite{HsuLin2009,Xu2010} for the interplay between graph theory and interconnection network design; and to \cite{DallyTowles2003} for an overview of interconnection networks and their implementation for distributed-memory multiprocessors. We implicitly refer to these references throughout.

\section{The dual-port server-centric DCN design space}
\label{sec:preliminaries}

A dual-port server-centric DCN can be built from: COTS servers, each with (at most) two network interface card (NIC) ports; dumb
``crossbar'' switches; and the cables that connect these hardware
components together.  We define the capability of a dumb crossbar-switch (henceforth referred to as a switch) as being able to forward an incoming packet to a single port requested in the packet header and to handle all such traffic in a non-blocking manner.  Such a switch only ever receives packets destined for servers directly attached to it and handles these requests by retrieving addresses from a very small forwarding table.  Consequently, it is never the case that two switches in the network are directly connected by a cable.  

We take a (primarily) mathematical view of datacenters in order to
systematically identify potential DCN topologies, and we abstract a DCN as an undirected graph so as to model only the major hardware components; namely, the servers and switches are abstracted as server-nodes and switch-nodes, respectively, and the interconnecting cables as edges or links. As our server-centric DCNs are dual-port, our graphs are such that every server-node has degree at most $2$ and the switch-nodes form an independent set in the graph. 

\subsection{Designing DCNs with good networking properties}
\label{sec:dcns-with-good-networking}

There are well-established performance metrics for DCNs and their routing
algorithms so that we might evaluate properties such as network throughput, latency, load balancing
capability, fault-tolerance, and cost to build (we'll return to these metrics later when we outline our methodology, in Section~\ref{sec:setup}, and undertake our empirical analysis, in Section~\ref{sec:evaluation}).  Networks
that perform well with respect to these or related metrics are said to
have \emph{good networking properties}.  Maintaining a diverse pool of
potential DCN topologies with good networking properties gives DCN
designers greater flexibility.  There is already such a pool of interconnection networks, developed over the past $50$ years or so, and it is precisely from here that the switch-centric DCN
fabrics of layer-2 switch-nodes in fat-trees and related topologies have been adapted
(see, \emph{e.g.}, \cite{Al-FaresLoukissasVahdat2008,Leiserson1985}).

Adapting interconnection networks to build server-centric DCNs, which necessarily have a more sophisticated arrangement of server-nodes and switch-nodes, however, is more complicated.
For example, BCube \cite{GuoLuLi2009} is built from a generalized
hypercube (see Definition~\ref{def:GQ}) by replacing the edges of
certain cliques, each with a switch-node connected to the nodes of
the clique.  In doing so, BCube inherits well-known routing algorithms
for generalized hypercubes, as well as mean-distance, fault-tolerance,
and other good networking properties.  DPillar \cite{LiaoYinYin2012}, which we discuss in
detail in Section \ref{sec:DPillar}, is built in a similar
manner from a wrapped butterfly network (see, \emph{e.g.}, \cite{Leighton1992}) by replacing bicliques\footnote{A biclique is a graph formed from two independent sets so that every node in one independent set is joined to every node in the other independent set.} with switch-nodes.  The presence
of these cliques and bicliques are inherent in the definitions of
generalized hypercubes and wrapped butterfly networks, respectively,
but are not properties of interconnection networks in
general. Furthermore, the dual-port property of DPillar is not by
design of the construction, but is a result of the fact that each node
in a wrapped butterfly is in exactly two maximal bicliques.

In order to effectively capitalise on a wide range of interconnection networks,
for the purpose of server-centric DCN design, we must devise
new generic construction methods, similar to those used to construct BCube and DPillar but that do not
impose such severe structural requirements on the interconnection
network used as the starting point.

\subsection{Related work}
\label{sec:dual-port-networks}

We briefly survey the origins of the dual-port server-centric DCNs
proposed thus far within the literature
\cite{LiGuoWu2011,GuoChenLi2013,LiGuoYang2014,LiaoYinYin2012,LiWu2015a},
referring the reader to the original publications for definitions of
topologies not given below. FiConn
\cite{LiGuoWu2011} is an adaptation of DCell and is unrelated to any
particular interconnection network. DPillar's origins \cite{LiaoYinYin2012} were discussed
above.  The topologies HCN and BCN
\cite{GuoChenLi2013} are built by combining a 2-level DCell with
another network, later discovered to be related to WK-recursive
interconnection networks \cite{Stewart2014,Della-VecchiaSanges1988}.  BCCC \cite{LiGuoYang2014} is a tailored construction
related to BCube and based on cube-connected-cycles and generalized
hypercubes.  Finally, SWKautz, SWCube, and SWdBruijn
\cite{LiWu2015a} employ a subdivision rule similar to ours, but the
focus in \cite{LiWu2015a} is not on the (generic) benefits of subdividing
interconnection networks as much as it is on the evaluation of those
two particular network topologies.

In Section~\ref{sec:evaluation} we compare an instantiation of our construction, namely the dual-port server-centric DCN GQ$^\star$, to FiConn and
DPillar.  The rationale for using these DCNs in our evaluation is
that they are good representatives of the spectrum of dual-port server-centric DCNs mentioned above: FiConn is a good
example of a DCN that includes both server-node-to-server-node and
server-node-to-switch-node connections and is somewhat unstructured, whereas
DPillar is server-node symmetric\footnote{Meaning that for every
  pair $(u,v)$ of server-nodes, there is an automorphism of the network
  topology that maps $u$ to $v$.} \cite{EricksonKiasariNavaridas2015f} and features only server-node-to-switch-node
connections.  In addition, FiConn is arguably unrelated to any
previously known interconnection network topology,
whilst DPillar is built from, and inherits some of the properties
of, the wrapped butterfly network.  Various other dual-port server-centric
DCNs lie somewhere between these two extremes.  Notice that neither
FiConn nor DPillar can be described as an instance of our
generalised construction: FiConn has some server-nodes whose only
connection is to a solitary switch-node, and in DPillar each server-node is
connected only to 2 switch-nodes. We now describe the constructions of the
DCNs FiConn and DPillar.

\subsection{The construction of FiConn}
\label{FiConn}

\begin{figure}[ht]
\centering
\includegraphics[width=.65\textwidth]{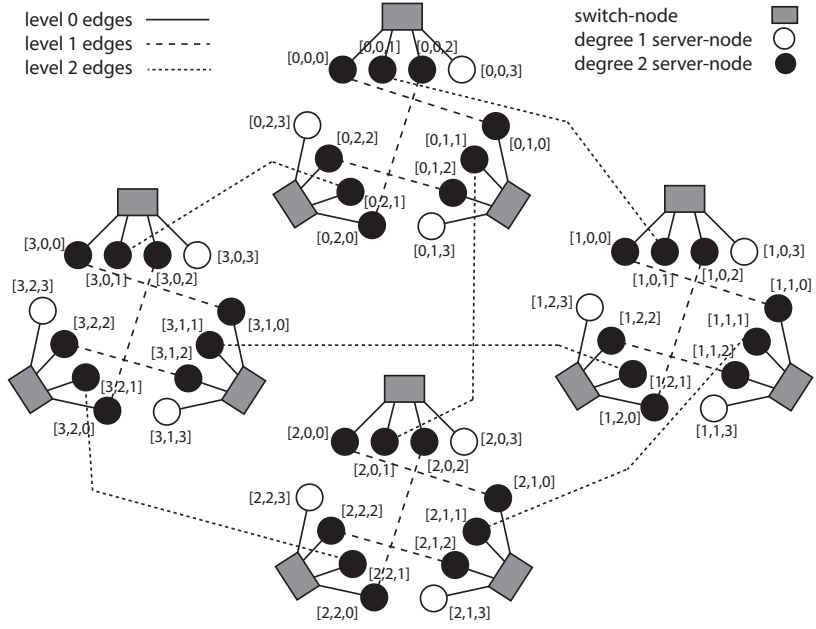}
\caption{A visualisation of FiConn$_{2,4}$.}
\label{Ficonnpic}
\end{figure}

We start with FiConn, the first dual-port server-centric DCN to be proposed and, consequently, typically considered the baseline such DCN.  For any even $n\geq 2$ and $k\geq 0$, FiConn$_{k,n}$ \cite{LiGuoWu2011} is a recursively-defined DCN where $k$ denotes the level of the recursive construction and $n$ the number of server-nodes that are directly connected to a switch-node (so, all
switch-nodes have degree $n$).  FiConn$_{0,n}$ consists of $n$ server-nodes and one switch-node, to which all the server-nodes are connected. Suppose that FiConn$_{k,n}$ has $b$ server-nodes of degree $1$ ($b=n$ when $k=0$; moreover, no matter what the value of $k$, $b$ can always be shown to be even). In order to build FiConn$_{k+1,n}$, we take $\frac{b}{2}+1$ copies of FiConn$_{k,n}$ and for every copy we connect one server-node of degree $1$ to each of the other $\frac{b}{2}$ copies
(these additional links are called level $k$ links).  The actual
construction of which server-node is connected to which is detailed precisely in \cite{LiGuoWu2011} (FiConn$_{2,4}$, as constructed in \cite{LiGuoWu2011}, can be visualised in
\Cref{Ficonnpic}); in particular,
there is a well-defined naming scheme where server-nodes of
FiConn$_{k,n}$ are named as specific $k$-tuples of integers. In fact, although it is not made clear in \cite{LiGuoWu2011}, there is a multitude of connection schemes realising different versions of FiConn. Note that all of the DCNs we consider in this paper come in parameterized families; so, when we say ``the DCN FiConn'', what we really mean is ``the family of DCNs $\{\mbox{FiConn}_{k,n}: k\geq 0, n \geq 2 \mbox{ is even}\}$''.

In \cite{LiGuoWu2011}, two routing algorithms are supplied: \emph{TOR\/} (traffic-oblivious routing) and \emph{TAR\/} (traffic-aware routing). \emph{TAR\/} is intended as a routing algorithm that dynamically adapts routes given changing traffic conditions (it was remarked in \cite{LiGuoWu2011} that it could be adapted to deal with link or port faults).

\subsection{The construction of DPillar}\label{sec:DPillar}

\begin{figure}[ht]
 \centering
 \includegraphics[width=0.65\textwidth]{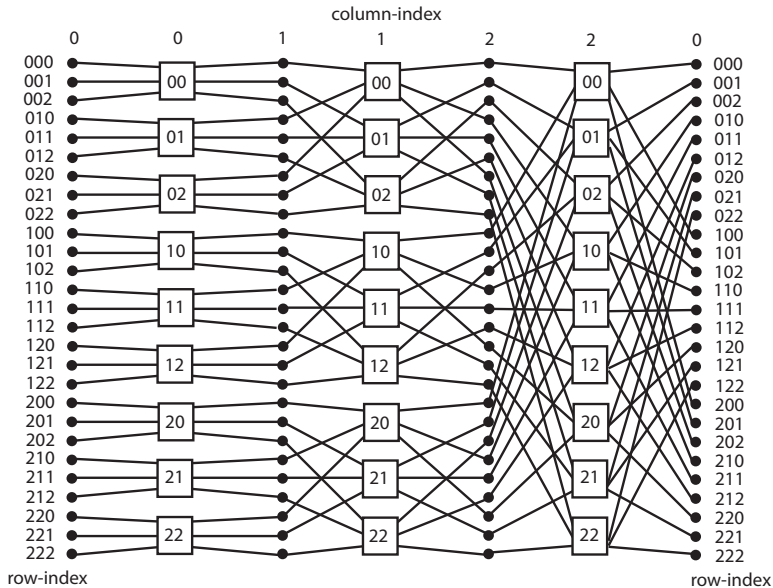}
 \caption{A visualization of DPillar$_{3,6}$. Squares represent switch-nodes, whereas dots represent server-nodes. For the sake of simplicity, the left-most and the right-most server-columns are the same (server-column 0).}
 \label{fig:DPillar}
 \end{figure}

The DCN DPillar$_{k,n}$ \cite{LiaoYinYin2012}, where $n\geq 2$ is even and $k\geq 2$, is such that $n$ denotes the
number of ports of a switch-node and $k$ denotes the level of
the recursive construction; it can be imagined as $k$ columns
of server-nodes and $k$ columns of switch-nodes, arranged alternately on the surface of a cylindrical pillar (see as an
example DPillar$_{3,6}$ in \Cref{fig:DPillar}). Each server-node in some server-column is adjacent to $2$ switch-nodes, in different adjacent switch-columns. Each server-column has $(\frac{n}{2})^k$ server-nodes, named as $\{0,1,\ldots,\frac{n}{2}-1\}^k$, whereas each switch-column has $(\frac{n}{2})^{k-1}$ switch-nodes, named as $\{0,1,\ldots,\frac{n}{2}-1\}^{k-1}$. We remark that in the literature, our DPillar$_{k,n}$ is usually referred to as DPillar$_{n,k}$. However, we have adopted our notation so as to be consistent with other descriptions of DCNs.

Fix $c\in\{0,1,\ldots,k-1\}$.  The server-nodes in server-columns
$c,c+1\in\{0,1,\ldots,k-1\}$ (with addition modulo $k$) are arranged into $(\frac{n}{2})^{k-1}$ groups of $n$ server-nodes so that in server-columns $c$ and $c+1$, the server-nodes in group
$(u_{k-1},\ldots,u_{c+1},u_{c-1},\ldots,u_0)\in\{0,1,\ldots,\frac{n}{2}-1\}^{k-1}$ are the server-nodes named
$\{(u_{k-1},\ldots,u_{c+1},i,u_{c-1},\ldots,u_0):
i\in\{0,1,\ldots,\frac{n}{2}-1\}\}$.
The adjacencies between switch-nodes and server-nodes are such that any server-node in group $(u_{k-1},\ldots,u_{c+1},u_{c-1},\ldots,u_0)$ in server-columns $c$ and $c+1$ is adjacent to the switch-node of name $(u_{k-1},\ldots,u_{c+1},u_{c-1},\ldots,u_0)$ in switch-column $c$.

In \cite{LiaoYinYin2012}, two routing algorithms are supplied: \emph{DPillarSP\/} and \emph{DPillarMP\/}. The former is a single-path routing algorithm and the latter is a multi-path routing algorithm.

While all of the dual-port server-centric DCNs from the literature have merit, it is clear that a generic method of transforming interconnection networks into dual-port server-centric DCNs has not previously been proposed and analysed. Having justified the value in studying the dual-port restriction, and having discussed the benefits of tapping into a large pool of potentially useful topologies, we proceed by presenting our generic construction in detail in the next section.

\section{Stellar DCNs: a new generic construction}
\label{sec:generic}

In this section we present our generic method of transforming
interconnection networks into potential dual-port server-centric DCNs.  We then describe how networking properties of the DCN, including routing algorithms, are inherited from the interconnection network, and go on to identify a preliminary pool of interconnection networks that particularly suit the stellar transformation. Next, we apply our stellar transformation in detail to generalized hypercubes as a prelude to an extensive empirical evaluation in Sections~\ref{sec:setup} and \ref{sec:evaluation}. The key aspects of our stellar construction are its topological simplicity, its universal applicability, and the tight relationship between the interconnection network and the resulting stellar DCN (in a practical networking sense). While we present our stellar construction within a graph-theoretic framework, we end this section by briefly discussing concrete networking aspects of our construction in relation to implementation. We remind the reader that we use \cite{DallyTowles2003,HsuLin2009,Xu2010} as our sources of information for the definitions and the networking properties of the families of interconnection networks mentioned below; we use these sources implicitly and only cite other sources when pertinent.

\subsection{The stellar construction}\label{sec:stellarcons}

An \emph{interconnection network\/} is an undirected  graph together with associated routing algorithms, packet-forwarding methodologies, fault-tolerance processes, and so on. However, it suffices for us to abstract an interconnection network as simply a graph $G=(V,E)$ that is undirected and without self-loops.

Let $G=(V,E)$ be any non-trivial connected graph, which we call the
\emph{base graph} of our construction. The \emph{stellar} DCN
$G^\ast$ is obtained from $G$ by placing $2$ server-nodes on each link
of $G$ and identifying the original nodes of $G$ as switch-nodes (see
\Cref{fig:knkstar}). We use the term ``stellar'' as we essentially
replace every node of $G$ and its incident links with a ``star''
subnetwork consisting of a hub switch-node and adjacent
server-nodes. Clearly, $G^\ast$ has $2|E|$ server-nodes and $|V|$
switch-nodes, with the degree of every server-node being $2$ and the
degree of every switch-node being identical to the degree of the
corresponding node in $G$.

We propose placing $2$ server-nodes on every link of $G$ so as to
ensure: uniformity, in that every server-node is adjacent to exactly
$1$ server-node and exactly $1$ switch-node (uniformity, and its stronger counterpart symmetry, are widely accepted as beneficial properties in general interconnection networks); that there are no links
incident only with switch-nodes (as this would violate the server-centric restriction, discussed in
Section~\ref{sec:preliminaries}); and that we can incorporate as many
server-nodes as needed within the construction (subject to the other
conditions). In fact, \emph{any} DCN in which every server-node is adjacent to exactly $1$
server-node and $1$ switch-node and where every switch-node is only
adjacent to server-nodes can be realised as a stellar DCN
$G^\ast$, for some base graph $G$.  In addition, the stellar transformation can be
applied to \emph{any} (non-trivial connected) base graph; that is, the
transformation does not rely on any non-trivial structural properties
of the base graph.

\subsection{Topological properties of stellar DCNs}
\label{sec:properties}

The principal decision that must be taken when constructing a
stellar DCN is in choosing an appropriate base graph $G$.  The
good networking properties discussed in
Section~\ref{sec:dcns-with-good-networking} are underpinned by several
graph-theoretic properties that are preserved under the stellar
transformation: for example, low diameter, high connectivity, and efficient routing
algorithms in the base graph $G$ translate more-or-less directly into
good (theoretical) networking properties of the stellar graph $G^\star$, as we now discuss.  The DCN
designer, having specific performance targets in mind, can use this
information to facilitate the selection of a base graph $G$ that meets
the requirements of the desired stellar DCN.

\subsubsection{Paths}

A useful aspect of our construction is as regards the transformation of paths in $G$ to paths in
$G^\star$.  As is usual in the analysis of server-centric DCNs (see,
e.g., \cite{GuoWuTan2008,GuoChenLi2013,GuoLuLi2009,LiGuoWu2011}), we
measure a server-node-to-server-node path $P$ by its
\emph{hop-length}, defined as one less than the number of server-nodes
in $P$.  Accordingly, we prefix other path-length-related measures
with hop-; for example, the hop-length of a shortest path joining two given server-nodes in $G^\star$ is the \emph{hop-distance} between the two server-nodes, and the
\emph{hop-diameter} of a server-centric DCN is the maximum over the hop-distances for every possible pair of server-nodes.  Let $G=(V,E)$
be a connected graph and let $u,v\in V$.  Let $u^\star$ and $v^\star$ be
the switch-nodes of $G^\star$ corresponding to $u$ and $v$, respectively.
Let $u'$ and $v'$ be server-node neighbours of $u^\star$ and $v^\star$,
respectively, in $G^\star$.  Each $(u,v)$-path $P$ in $G$, of length $m$, corresponds
uniquely to a $(u',v')$-path in $G^\star$ of hop-length $2m-1$, $2m$, or
$2m+1$. The details are straightforward.

\subsubsection{Path-based sub-structures}

The transformation of paths in $G$ to paths in $G^\star$ is the basis
for the transfer of potentially useful sub-structures in $G$ to $G^\star$ so as to yield good DCN
properties.  Any useful (path-based) sub-structure in $G$, such as a spanning tree,
a set of node-disjoint paths, or a Hamiltonian cycle, corresponds uniquely
to a closely related sub-structure in $G^\star$.  Swathes of research papers have uncovered
these sub-structures in interconnection networks, and the stellar
construction facilitates their usage in dual-port server-centric DCNs.
It is impossible to cover this entire topic here, but we describe how
a few of the more commonly sought-after sub-structures behave under the
stellar transformation.

\begin{figure}[ht]
      \centering
    \begin{subfigure}[b]{0.3\textwidth}
\centering 
      \def\svgwidth{.9\textwidth}  
      %
%% Creator: Inkscape inkscape 0.91, www.inkscape.org
%% Accompanies image file 'k33.eps' (pdf, eps, ps)
\begingroup%
  \makeatletter%
  \providecommand\color[2][]{%
    \errmessage{(Inkscape) Color is used for the text in Inkscape, but the package 'color.sty' is not loaded}%
    \renewcommand\color[2][]{}%
  }%
  \providecommand\transparent[1]{%
    \errmessage{(Inkscape) Transparency is used (non-zero) for the text in Inkscape, but the package 'transparent.sty' is not loaded}%
    \renewcommand\transparent[1]{}%
  }%
  \providecommand\rotatebox[2]{#2}%
  \ifx\svgwidth\undefined%
    \setlength{\unitlength}{194.400004bp}%
    \ifx\svgscale\undefined%
      \relax%
    \else%
      \setlength{\unitlength}{\unitlength * \real{\svgscale}}%
    \fi%
  \else%
    \setlength{\unitlength}{\svgwidth}%
  \fi%
  \global\let\svgwidth\undefined%
  \global\let\svgscale\undefined%
  \makeatother%
  \begin{picture}(1,1.00000004)%
    \put(0,0){\includegraphics[width=\unitlength]{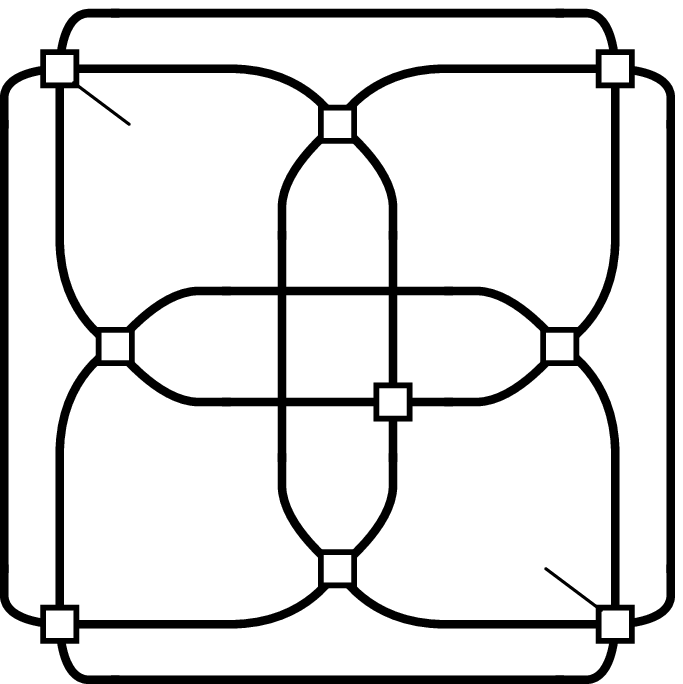}}%
    \put(0.20781895,0.79629629){\color[rgb]{0,0,0}\makebox(0,0)[lb]{\smash{$u$}}}%
    \put(0.75102881,0.17901229){\color[rgb]{0,0,0}\makebox(0,0)[lb]{\smash{$v$}}}%
  \end{picture}%
\endgroup%
       % \caption{The graph GQ\pms 3 3.}
    \end{subfigure}
     %\hfill %
    \begin{subfigure}[b]{0.3\textwidth}
      \centering
      \def\svgwidth{.9\textwidth}  
      %
%% Creator: Inkscape inkscape 0.91, www.inkscape.org
%% Accompanies image file 'k33star.eps' (pdf, eps, ps)
\begingroup%
  \makeatletter%
  \providecommand\color[2][]{%
    \errmessage{(Inkscape) Color is used for the text in Inkscape, but the package 'color.sty' is not loaded}%
    \renewcommand\color[2][]{}%
  }%
  \providecommand\transparent[1]{%
    \errmessage{(Inkscape) Transparency is used (non-zero) for the text in Inkscape, but the package 'transparent.sty' is not loaded}%
    \renewcommand\transparent[1]{}%
  }%
  \providecommand\rotatebox[2]{#2}%
  \ifx\svgwidth\undefined%
    \setlength{\unitlength}{198.35bp}%
    \ifx\svgscale\undefined%
      \relax%
    \else%
      \setlength{\unitlength}{\unitlength * \real{\svgscale}}%
    \fi%
  \else%
    \setlength{\unitlength}{\svgwidth}%
  \fi%
  \global\let\svgwidth\undefined%
  \global\let\svgscale\undefined%
  \makeatother%
  \begin{picture}(1,1.00000008)%
    \put(0,0){\includegraphics[width=\unitlength]{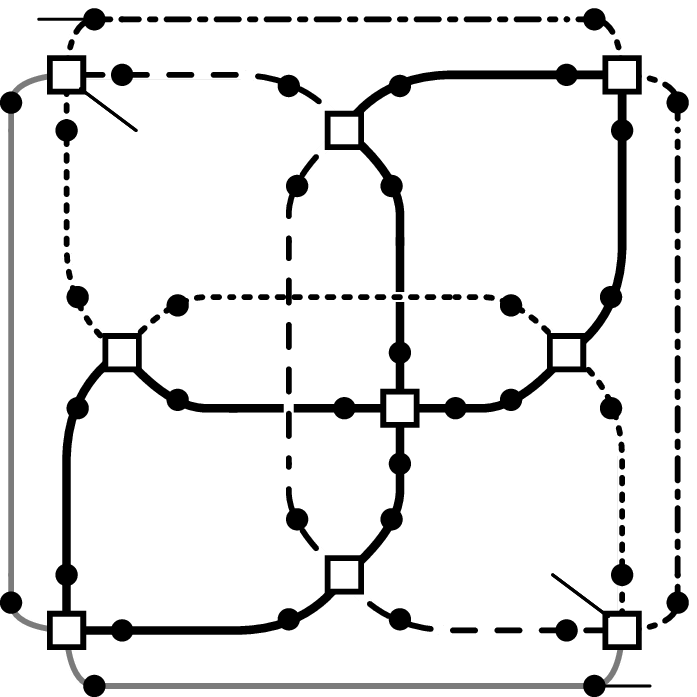}}%
    \put(0.21363751,0.79039578){\color[rgb]{0,0,0}\makebox(0,0)[lb]{\smash{$u^\star$}}}%
    \put(0.74602975,0.18540455){\color[rgb]{0,0,0}\makebox(0,0)[lb]{\smash{$v^\star$}}}%
    \put(0.00794051,0.93559367){\color[rgb]{0,0,0}\makebox(0,0)[lb]{\smash{$u'$}}}%
    \put(0.95172675,0.01600702){\color[rgb]{0,0,0}\makebox(0,0)[lb]{\smash{$v'$}}}%
    \put(0.21363751,0.79039578){\color[rgb]{0,0,0}\makebox(0,0)[lb]{\smash{$u$}}}%
    \put(0.74602975,0.18540455){\color[rgb]{0,0,0}\makebox(0,0)[lb]{\smash{$v$}}}%
  \end{picture}%
\endgroup%
     \end{subfigure}
    \caption{Transforming $4$ paths from $u$ to
      $v$ in $G$ (\emph{left\/}) into $4$ paths from $u'$ to $v'$ in $G^\star$ (\emph{right\/}).}
    \label{fig:knkstar}
\end{figure}

Foremost are internally node-disjoint paths, associated with fault-tolerance and load balancing.  As
the degree of any server-node in $G^\ast$ is $2$, one cannot hope to
obtain more than $2$ internally node-disjoint paths joining any $2$ distinct
server-nodes of $G^\ast$.  However, a set of $c$ internally node-disjoint
$(u,v)$-paths in $G$ corresponds uniquely to a set of $c$ internally
(server- and switch-) node-disjoint $(u^\star,v^\star)$-paths in $G^\star$, where
$u,v,u^\star,v^\star,u',$ and $v'$ are as defined above.  This provides a set
of $c$ $(u',v')$-paths in $G^\star$, called \emph{parallel paths},
that are internally node-disjoint apart from possibly $u^\star$ and
$v^\star$ (see \Cref{fig:knkstar}).  It is trivial to show that the minimum
number of parallel paths between any pair of server-nodes, not
connected to the same switch-node, in $G^\star$ is equal to the connectivity of $G$.

%\begin{figure}[ht]
%      \centering
%    \includegraphics[width=0.65\textwidth]{img/k33starnew.eps}
%    \caption{Transforming $4$ paths from $u$ to
%      $v$ in $G$ (\emph{left\/}) into $4$ paths from $u'$ to $v'$ in $G^\star$ (\emph{right\/}).}
%      \label{fig:knkstar}
%\end{figure}

By reasoning as above, it is easy to see that a set of $c$ edge-disjoint
$(u,v)$-paths in $G$ becomes a set of $c$ internally
server-node-disjoint $(u',v')$-paths in $G^\star$, with
$u,v,u^\star,v^\star,u',$ and $v'$ defined as above; we shall call these
\emph{server-parallel paths}. The implication is that as any two of these
paths share only the links $(u',u^\star)$ and
$(v^\star,v')$, a high value of $c$ may be leveraged to alleviate
network traffic congestion as well as fortify the network against
server-node failures.
%That is, the paths in $G^\star$ share at most four ports or
% their source and destination switch- and server-nodes.

On a more abstract level, consider any connected spanning sub-structure
$H$ of $G$, such as a Hamiltonian cycle or a spanning tree.  Let
$H^\star$ be the corresponding sub-structure in $G^\star$ (under the path-to-path mapping described above) and observe
that each edge of $G$ not contained in $H$ corresponds to two adjacent
server-nodes in $G^\star$ not contained in $H^\star$.  On the other
hand, every server-node not in $H^\star$ is exactly one hop away from
a server-node that is in $H^\star$; so within an additive factor of
one hop, $H^\star$ is just as ``useful'' in $G^\star$ as $H$ is in
$G$. In fact, if $H$ is a spanning tree in $G$ then we can extend $H^\star$ in $G^\star$ by augmenting it with pendant edges from switch-nodes so that what results is a spanning tree in $G^\star$ containing \emph{all\/} server-nodes of $G^\star$ (and not just those in the original $H^\star$). By the same principle, non-spanning sub-structures of $G$, such
as those used in one-to-many, many-to-many, and many-to-one
communication patterns, also translate to useful sub-structures in $G^\star$.

%The power of the stellar transformation becomes even more apparent when combining the above two properties,  namely the proximity of server-nodes to a spanning sub-structure $H^\star$ and the transformation of edge-disjointness in $G$ into server-node disjointness in $G^\star$. We shall illustrate this with spanning trees but the following holds in a more general setting.  Let $S$ and $T$ be a pair of \emph{totally independent spanning trees} in $G$; that is, $S$ and $T$ are edge-disjoint, and for any pair of nodes $u$ and $v$, the $(u,v)$-path in $S$ is internally vertex-disjoint from the $(u,v)$-path in $T$.  Totally independent spanning trees have applications in efficient fault-tolerant broadcasting and have been studied in several of the interconnection networks we mention below in Section~\ref{sec:suited} (\emph{e.g.}, see \cite{HasunumaMorisaka2012}).  Let $S^\star$ and $T^\star$ be the structures corresponding to $S$ and $T$, respectively, within $G^\star$, and let $u^\star,v^\star,u',v'$ be defined as above.  Our earlier discussion on spanning sub-structures yields that $S^\star$ and $T^\star$ are server-node-disjoint, and for any pair of server-nodes $(u',v')$, there is a pair of parallel $(u',v')$-paths, one through $S^\star$ and the other through $T^\star$.

We summarise the relationship between properties of $G$ and $G^\star$
that we have discussed so far in Table~\ref{tab:transformation} where corresponding properties for $G$ and $G^\star$ are detailed. It should now be apparent that the simplicity of our stellar transformation enables us to import good networking properties from our base graphs to our stellar DCNs where these properties are crucial to the efficacy of a DCN.

\begin{table}[ht]
  \centering
  \caption{Transformation of networking properties of a connected graph $G$}
  \label{tab:transformation}
  \begin{tabular}{lll}
    property&$G=(V,E)$&$G^\star$\\
    \hline
    nodes/nodes& $|V|$ & $|V|$ switch-nodes\\
&& $2|E|$ server-nodes\\
    node degree/switch-node degree & $d$& $d$\\  
    edges/links& $|E|$ & $3|E|$ (bidirectional)\\
    path-length/hop-length& $x$ & $2x-1\le\cdot \le 2x+1$\\
    diameter/hop-diameter& $D$ & $2D-1,2D,$ or $2D+1$\\
    internally-disjoint paths/parallel paths&$\kappa$ & $\kappa$\\
    edge-disjoint paths/server-parallel paths &$\gamma$  & $\gamma$
  \end{tabular}
\end{table}

We close this sub-section with a brief discussion of the transferral of routing algorithms under the stellar transformation. A routing algorithm for an interconnection network $G$ is effectively
concerned with an efficient computation over some communication
sub-structures.  For example, in the case of unicast routing from $u$ to $v$, we may
compute one or more $(u,v)$-paths (and route packets over them), or
for a broadcast we may compute one or more spanning trees.  Routing algorithms can be executed at the source
node or in a distributed fashion, and they can be deterministic or
non-deterministic; whatsoever the process, the resulting output is a
communication sub-structure over which packets are sent from node to node.  We discussed above the correspondence between
communication sub-structures in $G$ and those in $G^\star$; we
now observe that, in addition, any routing algorithm on $G$ can be
simulated on $G^\star$ with the same time complexity.  We leave the
details to the reader (but we will instantiate this later when we build the stellar DCNs GQ$^\star$).

\subsection{A pool of suitable base graphs}
\label{sec:suited}

So far, we have referred to an interconnection network as a solitary object. However, interconnection networks (almost always) come in families where there are parameters the values for which precisely delineate the family members. For example the hypercube $Q_n$ is parameterized by the degree $n$ of the nodes, and so really by ``the hypercube $Q_n$'' we mean ``the family of hypercubes $\{Q_n:n=1,2,\ldots,\}$''. For the rest of this sub-section we will be precise and speak of families of interconnection networks as we need to focus on the parameters involved. To ease understanding, when there is more than $1$ parameter involved in some definition of a family of interconnection networks and these parameters appear as subscripts or in tuples in the denotation, we list parameters relating to the dimension of tuples or the depth of recursion first with parameters referring to the size of some component-set coming afterwards (we have done this so far with FiConn$_{k,n}$ and DPillar$_{k,n}$). We remark that this is sometimes at odds with standard practice in the literature. 

We validate our claim that many families of interconnection networks suit the stellar construction by highlighting several that, first, have parameters flexible enough to yield interconnection networks of varying and appropriate size and degree, and, second, are known to possess good networking properties. The first goal is to identify families of interconnection networks that have suitable combinations of degree and size, bearing in mind that today's DCN COTS switches have up to tens of ports, with 48 being typical, while conceivable (but not necessarily in production) sizes of DCNs range from tens of server-nodes up to, perhaps, $5$ million in the near future. An illustration of a family of interconnection networks \emph{lacking\/} this flexibility is the family of hypercubes, where the hypercube $Q_n$ necessarily has $2^n$ nodes when the degree is fixed at $n$; this translates to a stellar DCN with $n$-port switch-nodes and, necessarily, $n2^n$ server-nodes. As such, there is a lack of flexibility, in terms of the possible numbers of server-nodes, and if we were to build our stellar DCNs using commodity switches with 48 ports then we would have to have $48\times 2^{48}$ servers which is clearly impossible. Another illustration of a family of interconnection networks lacking flexibility is the family of cube-connected cycles $\{CCC(n): n\geq 3\}$, where $CCC(n)$ is obtained from a hypercube $Q_n$ via a transformation similar to our stellar transformation: $2$ new nodes are placed on each edge; the new nodes adajcent to some old node are joined (systematically) in a cycle of length $n$; and the old nodes, and any adjacent edges, are removed. So, $CCC(n)$ is regular of degree $3$ and consequently unsuitable for our stellar transformation.

We now look at some families of interconnection networks that \emph{are\/} suitable for our stellar transformation. It is too much to list all of the good networking properties of the interconnection networks discussed below. However, it should be remembered that, from above, any path, path-based sub-structure, and routing algorithm is immediately inherited by the stellar DCN; consequently, we focus on the flexibility of the parameterized definition in what follows and refer the reader to other sources (including \cite{DallyTowles2003,HsuLin2009,Xu2010}) for more details as regards good networking properties. Besides: the fact that these families of interconnection networks have featured so strongly within the research literature is testament to their good networking properties. Also, the families of interconnection networks mentioned below are simply illustrations of interconnection networks for which our stellar transformation has potential and there are many others not mentioned.

Tori (also known as toroidal meshes) have been widely studied as interconnection networks; indeed, tori form the interconnection networks of a range of distributed-memory multiprocessor computers (see, \emph{e.g.},\cite{DallyTowles2003}). The uniform version of a torus is the $n$-ary $k$-cube $Q_{k,n}$, where $k\geq 1$ and $n\geq 3$, whose node-set is $\{0,1,\ldots,n-1\}^k$ and where there is an edge joining two nodes if, and only if, the nodes differ in exactly one component and the values in this component differ by $1$ modulo $n$; hence, $Q_{k,n}$ has $n^k$ nodes and $kn^k$ edges, and every node has degree $2k$. There is some, though limited, scope for using $n$-ary $k$-cubes in our stellar construction. For example, if we use switch-nodes with $16$ ports to build our DCN then this means that $k=8$; choosing $n=3$, $4$, or $5$ results in our stellar DCN having $104,976$ server-nodes, $1,048,576$ server-nodes, or $6,250,000$ server-nodes, respectively. We get more variation if we allow the sets of values in different components to differ; that is, we use mixed-radix tori. However, it is not really feasible to use switch-nodes with more than $16$ ports in our stellar construction.

Circulant graphs have been studied extensively in a networking
context, where they are often called multi-loop networks.  Let $S$ be a set of integers, called \emph{jumps}, with $1\le s \le \flr{N}{2}$, for each $s\in S$, and where $N\ge 2$.  A circulant $G(N;S)$ has node set $\set{0,1,\ldots, N-1}$, where node $i$ is connected to nodes $i\pm s \pmod N$, for each $s\in S$.  A circulant has $N$ nodes and at most $N|S|$ edges, and the degree of every node is approximately $2|S|$ (depending upon the relative values of $N$ and the integers in $S$); consequently, the parameters provide significant general flexibility. Illustrations of good networking properties of circulants can be found in, for example, \cite{CaiHavasMans1999,Hwang2003,Monakhova2012}.

The wrapped butterfly network $BF(k,n)$ can be obtained from DPillar$_{k,n}$ by replacing all switch-nodes with bicliques (joining server-nodes in adjacent columns); consequently, $B(k,n)$ has $k(\frac{n}{2})^k$ nodes and $k(\frac{n}{2})^{k+1}$ edges, and each node has degree $n$. Thus, by varying $k$ and $n$, there is reasonable scope for flexibility in terms of the sizes of stellar DCNs. Illustrations of the good networking properties of wrapped butterfly networks can be found in, for example, \cite{FuChau1998,TouzeneDayMonien2005}. Note that transforming a wrapped butterfly network to obtain DPillar is different to transforming it according to the stellar transformation; the two resulting DCNs are combinatorially very distinct.

The de Bruijn digraph $dB(k,n)$, where $k\geq 1$ and $n\geq 2$ is even, is a graph with node-set $\{0,1,\ldots,\frac{n}{2}-1\}^k$.  There is a directed edge from $(s_{k-1},s_{k-2},\ldots,s_0)$ to $(s_{k-2},s_{k-3},\ldots,s_0,\alpha)$, for each $\alpha\in\{0,1,\ldots,\frac{n}{2}-1\}$. Undirected de Bruijn graphs are obtained by regarding all directed edges as undirected and removing self-loops and multiple edges; such graphs are not regular but nearly so, with most of the $(\frac{n}{2})^k$ nodes having degree $n$ although some have degree $n-1$ or $n-2$. Consequently, by varying the values of $k$ and $n$, there is good flexibility in terms of the sizes of stellar DCNs. Illustrations of the good networking properties of de Bruijn graphs can be found in, for example, \cite{PradhanReddy1982,EsfahanianHakimi85}. Note that de Bruijn graphs have been studied as server-centric DCNs in \cite{PopaRatnasamyIannaccone2010} but these DCNs are not dual-port.

The arrangement graph $A_{k,n}$, where $n\geq 2$ and $1\leq k\leq n-1$, has node-set $\{(s_{k-1},s_{k-2},\ldots,s_1,s_0): s_i \in \{0,1,\ldots,n-1\}, s_i\neq s_j, i,j = 0,1, \ldots,k-1\}$. There is an edge joining two nodes if, and only if, the nodes are identical in all but one component. Hence, the arrangement graph $A_{k,n}$ has $\frac{n!}{(n-k)!}$ nodes and $\frac{k(n-k)n!}{2(n-k)!}$ edges, and is regular of degree $k(n-k)$.  The family of arrangement graphs includes the well-known star graphs as a sub-family, and there is clearly 
considerable flexibility in their degree and size.  

\subsection{The stellar DCNs GQ$^\star$}\label{sec:GQS}

Having hinted that there are various families of interconnection networks to which our stellar transformation might sensibly be applied, we now apply the stellar transformation to one specific family in detail: the family of generalized hypercubes \cite{BhuyanAgrawal1984}. We provide below more details as regards the topological properties of and routing algorithms for generalized hypercubes as we will use these properties and algorithms in our experiments in Sections~\ref{sec:setup} and~\ref{sec:evaluation}. We choose generalized hypercubes because of their flexibility as regards the stellar construction, their good networking properties, and the fact that they have already featured in DCN design as templates for BCube.

\begin{definition}
  \label{def:GQ}
  The \emph{generalized hypercube\/} $GQ_{k,n}$, where $k\geq 1$ and $n\geq 2$, has node-set $\{0,1,\ldots,n-1\}^k$ and there is an edge joining two nodes if, and only if, the names of the two nodes differ in exactly one component.
\end{definition}

Consequently, $GQ_{k,n}$ has $n^k$ nodes and $\frac{1}{2}k(n-1)n^k$ edges, and is regular of degree $k(n-1)$. Also, $GQ_{k,n}$ has diameter $k$ and connectivity $k(n-1)$. Hence, GQ$_{k,n}^\star$ has hop-diameter $2k+1$ and there are $k(n-1)$ parallel paths between any two distinct server-nodes. 

Suppose that we wished to use 48-port switch-nodes (and utilize all switch-ports) in a stellar DCN built from $GQ_{k,n}$. We might choose $(k,n)$ as $(2,25)$, $(3,17)$, or $(4,13)$ with the result that the number of server-nodes is $30,000$ for GQ$^\ast_{2,25}$, $235,824$ for GQ$^\ast_{3,17}$, or $1,370,928$ for GQ$^\ast_{4,13}$, respectively (of course, we can vary this number of server-nodes if we do not use all switch-ports or if we use switch-nodes with less than 48 ports). 

The stellar construction allows us to transform existing routing
algorithms for the base graph $GQ_{k,n}$ into routing algorithms for GQ$^\star_{k,n}$.  We describe this process using the routing algorithms for $GQ_{k,n}$ surveyed in
\cite{YoungYalamanchili1991}. Let $u = (u_{k-1}, u_{k-2}, \ldots , u_0)$ and $v = (v_{k-1}, v_{k-2},
\ldots, v_0)$ be two distinct nodes of $GQ_{k,n}$. The basic
routing algorithm for $GQ_{k,n}$ is \emph{dimension-order\/} (or
\emph{e-cube\/}) routing where the path from $u$ to $v$ is
constructed by sequentially replacing each $u_i$ by $v_i$, for some
predetermined ordering of the coordinates, say $i=0,1\ldots, k-1$.  As we mentioned above, 
dimension-order routing translates into a shortest-path routing
algorithm for GQ$^\star_{k,n}$ with unchanged time complexity, namely $O(k)$.

We introduce a fault-tolerant mechanism called
\emph{intra-dimensional\/} routing by allowing the path to replace $u_i$ by $v_i$ in two steps, using a \emph{local proxy\/}, rather than in one
step, as described in dimension-order routing.  Suppose, for example,
that one of the edges in the dimension order route from $u$
to $v$ is faulty; say, the one from $u=(u_{k-1},u_{k-2},\ldots, u_1,u_0)$ to $x=(u_{k-1},u_{k-2},\ldots, u_1,v_0)$ (assuming that $u_0$ and $v_0$ are distinct).  In this case we can
try to hop from $u$ to $(u_{k-1},u_{k-2},\ldots,u_1,x_0)$, where
$u_0\neq x_0\neq v_0$, and then to $x$.

\emph{Inter-dimensional\/} routing is a routing algorithm that extends intra-dimensional routing so that if intra-dimensional routing fails, because a local proxy within a specific dimension cannot be used to re-route round a faulty link, an alternative dimension is chosen. For example, suppose that in $GQ_{k,n}$ intra-dimensional routing has successfully built a route over dimensions $1$ and $2$ but has failed to re-route via a local proxy
in dimension $3$. We might try and build the route instead over dimension $4$ and then return and try again with dimension $3$. Note that if a non-trivial path extension was made in dimension $4$ then this yields an entirely different locality within GQ$_{k,n}$ when trying again over dimension $3$.

However, in our upcoming experiments we implement the most extensive fault-tolerant, inter-dimensional
routing algorithm possible, called \emph{GQ$^\star$-routing\/}, for the stellar DCN GQ$^\star_{k,n}$,
whereby we perform a depth-first search of the dimensions and we use
intra-dimensional routing to cross each dimension wherever necessary
(and possible).  In addition, if \emph{GQ$^\star$-routing\/} fails to route
directly in this fashion then it attempts four more times to route (as
above) from the source to a randomly chosen server-node, and from there to
the destination.  We have chosen to make this extensive search of
possible routes in order to test the maximum capability of
\emph{GQ$^\star$-routing\/}; however, we expect that in practice the best
performance will be obtained by limiting the search in order to avoid
certain worst-case scenarios. The precise implementation details of \emph{GQ$^\star$-routing\/} can be found in the software release of \snapflow\ \cite{TEF} (see Section~\ref{sec:setuptools}). Finally, it is easy to see that \emph{GQ$^\star$-routing\/} can be implemented as a  distributed algorithm if a small amount of extra header information is attached to a path-probing packet, similarly to the suggestion in \cite{LiGuoWu2011} for implementing \emph{TAR\/} (Traffic Aware Routing) in FiConn.

\subsection{Implementing stellar DCNs}
\label{sec:implementing}

Implementing the software suite from scratch would require a software
infrastructure that supports through-server end-to-end communications.
This could be implemented either on top of the transport layer (TCP)
so as to simplify development, since most network-level mechanisms
(congestion control, fault-tolerance, quality of service) would be
provided by the lower layers. Alternatively, it could be implemented
on top of the data-link layer to improve the performance, since a
lower protocol stack will result in the faster processing of packets.  The
latter would require a much higher implementation effort in order to
deal with congestion and reliability issues. At any rate, the design
and development of a software suite for server-centric DCNs is outside
the scope of this paper, but may be considered in the future.

\section{Methodology}
\label{sec:setup}

The good networking properties discussed in Section~\ref{sec:dcns-with-good-networking} guide our evaluation methodology; they are network throughput, latency, load balancing capability, fault-tolerance, and cost to build.  These properties are reflected through performance metrics, and in this section we explain how we use aggregate bottleneck throughput, distance metrics, connectivity, and paths and their congestion, combined with a selection of traffic patterns, in order to evaluate the performance of our DCNs and routing algorithms. In particular, we describe and justify the use of our simulation tool in Section~\ref{sec:setuptools}.

Our methodological framework is as follows. First, we take the position, similar to Popa \emph{et al.}
\cite{PopaRatnasamyIannaccone2010}, that the cost of a network is of fundamental importance.  No matter what purpose a network is intended for, the primary objective is to maximise the return on the cost of a DCN. While there are several elements that factor into the cost of a DCN, including operational costs, our concern is with the capital costs of purchasing and installing the components we are modelling: servers, switches, and cables. Having calculated these costs (in Section~\ref{sec:network-cost} below), where appropriate (in our evaluation in Section~\ref{sec:scalability}) we normalise with respect to cost and proceed by both quantitatively and qualitatively interpreting the resulting multi-dimensional metric-based comparison.  Subsequently, (from Section~\ref{sec:evaluate-fault-tolerance} onwards) we focus on $4$ carefully chosen DCNs, namely GQ$^\star_{3,10}$, GQ$^\star_{4,6}$, FiConn$_{2,24}$, and DPillar$_{4,18}$, and evaluate these DCNs in some detail. We have selected these DCNs as their properties are relevant to the construction of large-scale DCNs: they each have around $25,000$ server-nodes and use switch-nodes of around $24$ ports.  Table~\ref{table:DCNs} details some of their topological properties.

\begin{table}[ht]
\centering
\caption{Basic properties of the selected DCNs.}
\begin{tabular}{c|cccc}
        topology&GQ$^\star_{3,10}$&GQ$^\star_{4,6}$&FiConn$_{2,24}$&DPillar$_{4,18}$\\
        \hline
server-nodes & $27,000$ & $25,920$ & $24,648$ & $26,244$\\
switch-nodes & $1,000$ & $1,296$ & $1,027$ & $2,916$\\
switch-ports & $27$ & $20$ & $24$ & $18$\\
links & $81,000$ & $77,760$ & $67,782$ & $104,976$\\
diameter & $7$ & $9$ & $7$ & $7$\\
parallel paths & $27$ & $20$ & unknown & $9$ (see \cite{LiaoYinYin2012}) \\
%max flows & $99,359$ & $111,479$ & $121,367$ & $141,062$\\
%ABT & $7,336$ & $6,026$ & $5,005$ & $4,882$
\end{tabular}
\label{table:DCNs}
\end{table}

\subsection{Network cost}
\label{sec:network-cost}

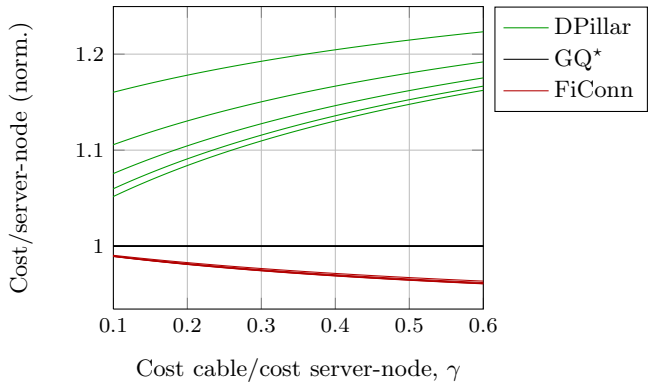
\begin{figure}[ht]
 \centering
%
% http://sourceforge.net/projects/pgfplots/
%\usepackage{pgfplots}
%\usetikzlibrary{patterns}
%\begin{document}
% Preamble: \pgfplotsset{width=7cm,compat=1.10} %\documentclass{article} %\usepackage{pgfplots} %\pgfplotsset{compat=1.5}

\begin{tikzpicture}
  \begin{axis}[width=0.5\linewidth,
    height=0.2\linewidth,
    small,
    sharp plot,
    xlabel={Cost cable/cost server-node, $\gamma$},%
    ylabel={Cost/server-node (norm.)}, %
    xmin=0.1,
    xmax=.6,
    legend entries={
      {DPillar},
      {GQ$^\star$},
      {FiConn},
    },
    legend style={cells={anchor=west},font=\small,},
    legend pos=outer north east,
    grid=major,
    thin,
    mark size=0pt,
    ]

\addplot[green!60!black]
   table{dat/cost/d_r0.01.dat};
   \addplot[black]
   table{dat/cost/gq_r0.01.dat};
\addplot[red!70!black]
   table{dat/cost/f_r0.01.dat};

   \addplot[green!60!black]
   table{dat/cost/d_r0.02.dat};
\addplot[green!60!black]
   table{dat/cost/d_r0.04.dat};
\addplot[green!60!black]
   table{dat/cost/d_r0.08.dat};
\addplot[green!60!black]
   table{dat/cost/d_r0.16.dat};

\addplot[black]
   table{dat/cost/gq_r0.02.dat};
\addplot[black]
   table{dat/cost/gq_r0.04.dat};
\addplot[black]
   table{dat/cost/gq_r0.08.dat};
\addplot[black]
   table{dat/cost/gq_r0.16.dat};

\addplot[red!70!black]
   table{dat/cost/f_r0.02.dat};
\addplot[red!70!black]
   table{dat/cost/f_r0.04.dat};
\addplot[red!70!black]
   table{dat/cost/f_r0.08.dat};
\addplot[red!70!black]
   table{dat/cost/f_r0.16.dat};

 \end{axis}
\end{tikzpicture}
 
%%% Local Variables: 
%%% mode: latex
%%% TeX-master: "GenDualPort"
%%% End: 
 \caption{The component costs per server-node of FiConn and
  DPillar, relative to that of GQ$^\star$, for $\rho\in
  \set{0.01,0.02,0.4,0.8,1.6}$.}
\label{plot:cost-r-g}
\end{figure}

We follow Popa \emph{et al.} \cite{PopaRatnasamyIannaccone2010} and assume that the cost of a switch is proportional to its radix (this is justified in \cite{PopaRatnasamyIannaccone2010} for switches of radix up to around $100$-$150$ ports). Let $c_s$ be the cost of a server, let $c_p$ be the cost of a switch-port, and let $c_c$ be the average cost of a cable.  We make the simplifying assumption that the average cost of a cable $c_c$ is uniform across DCNs with $N$ servers within the families GQ$^\star$, FiConn, and DPillar, and, furthermore, that the average cost of a cable connected only to servers is similar to that of a cable connected to a switch.  Thus, the cost of a DCN GQ$^\star$ with $N$ server-nodes is $N(c_p+c_c+c_s+\frac{c_c}{2})$; the cost of a DCN FiConn$_{k,n}$ with $N$ server-nodes is
$N(c_p+c_c+c_s+\frac{c_c}{2}-\frac{c_c}{2^{k+1}})$, since it contains $\frac{N}{2^k}$ server-nodes of degree 1 \cite{LiGuoWu2011}; and the cost of a DCN DPillar with $N$ server-nodes is $N(2(c_p+c_c)+c_s)$.  Next, we express $c_p=\rho c_s$ and $c_c=\gamma c_s$  so that the costs per server-node become $Nc_s(\rho+\gamma + 1 + \frac{\gamma}{2})$,
$Nc_s(\rho+\gamma+1+\frac{\gamma}{2}-\frac{\gamma}{2^{k+1}})$, and
$Nc_s(2(\rho+\gamma) + 1)$, respectively. A rough estimate is that realistic values for $\rho$ lie in the range $[0.01,0.1]$, and that realistic values for $\gamma$ lie in the range $[0.01,0.6]$; we choose the ranges conservatively because there is great variation in the cost of components, \emph{e.g.}, between copper and optical cables, as well as how we account for the labour involved in installing them.  Consequently, we normalise with respect to the aggregated component cost per server-node in GQ$^\star$, letting $c_s(\rho +\gamma + 1 + \frac{\gamma}{2})=1$, and plot component costs per server-node against $\gamma$ in \Cref{plot:cost-r-g}, for the representative selection $\rho\in\set{0.01,0.02,0.4,0.8,1.6}$ (in \Cref{plot:cost-r-g}, there is one graph for each DCN and for each of the 5 values for $\rho$, with the 5 graphs corresponding to FiConn being almost indistinguishable from one another).  The upshot is that the higher the value for $\rho$, the higher the cost of DPillar, and for the specific choices of $\rho$ and $\gamma$ mentioned above, DPillar could be up to 20\% more expensive and FiConn around 4\% less expensive than GQ$^\star$ when all DCNs have the same number of server-nodes. Perhaps the most realistic values of $\rho$ and $\gamma$, however, yield a DPillar that is only about 10\% more expensive and FiConn that is only about 2\% less expensive.

\subsection{Hop-distance metrics}
\label{sec:distance}
The number of servers a packet flow needs to travel through
significantly affects the flow's latency.  In addition, for each
server on the path, the compute and memory overheads are impacted
upon: in a server-centric DCN (with currently available COTS
hardware), the whole of the protocol stack, up to the application
level, needs to be processed at each server which can make message
transmission noticeably slower than in a switch-centric network where lower layers of the protocol stack are employed and use optimised implementations.

The paths over which flows travel are computed by routing algorithms, and it may not be the case that shortest-paths are achievable by available routing algorithms and without global fault-knowledge; large-scale networks like DCNs are typically restricted to routing algorithms that use only local knowledge of fault locations. As such, the performance of the routing algorithm is perhaps more important than the hop-diameter or mean hop-distance of the topology itself.  Therefore, we use distance-related metrics that reveal the performance of the topology and the routing algorithm combined, namely \emph{routed hop-diameter\/} and \emph{routed mean hop-distance\/}, as well as for the topology alone (where appropriate), namely hop-diameter and mean hop-distance (see Section~\ref{sec:dcns-with-good-networking}).  This allows us to (more realistically) assess both the potential of the topologies and the actual performance that can be extracted from them when implemented with currently available routing algorithms.

\subsection{Aggregate bottleneck throughput}
\label{sec:abt}
The \emph{aggregate bottleneck throughput\/} (\emph{ABT\/}) is a metric introduced in \cite{GuoLuLi2009} and is of primary interest to DCN designers due to its suitability for evaluating the worst-case throughput in the all-to-all traffic pattern, which is extremely significant in the context of DCNs (see Section~\ref{sec:traffic-patterns}).  The reasoning behind ABT is that the performance of an all-to-all operation is limited by its slowest flow, \emph{i.e.}, the flow with the lowest throughput.  The ABT is defined as the total number of flows times the throughput of the \emph{bottleneck flow\/}; that is, the link sustaining the most flows.  Formally, the ABT of a network of size $N$
is equal to $\frac{N(N-1)b}{F}$, where $F$ is the number of flows in the bottleneck link and $b$ is the bandwidth of a link.

In our experiments, the bottleneck flow is determined experimentally using the implementations of actual routing algorithms; this is atypical of ABT calculations (\emph{e.g.\/}, see \cite{LiuMuppalaVeeraraghavan2013}), where ordinarily shortest-paths are used, but our approach facilitates a more realistic evaluation.  We measure ABT using \emph{GQ$^\star$-routing\/} for GQ$^\star$, \emph{TOR\/} for FiConn, and \emph{DPillarSP\/} for \emph{DPillar\/}, assuming $N(N - 1)$ flows and a bandwidth of $1$ unit per directional link, where $N$ is the number of server-nodes.  Since datacenters are most commonly used as a stream processing platform, and are therefore bandwidth limited, this is an extremely important performance metric in the context of DCNs.  Given that ABT is only defined in the context of all-to-all communications, for other traffic patterns we focus on the number of flows in the bottleneck as an indicator of congestion propensity.

We should explain our choice of routing algorithm in FiConn and DPillar as regards our ABT analysis. In \cite{LiGuoWu2011}, it was shown that \emph{TOR\/} yields better performance for all-to-all traffic patterns than \emph{TAR\/}. In \cite{LiaoYinYin2012}, the all-to-all analysis (actually, it is a many all-to-all analysis) showed that \emph{DPillarSP\/} performs better than \emph{DPillarMP\/}. We have chosen \emph{TOR\/} and \emph{DPillarSP\/} so as not to disadvantage FiConn and DPillar when we compare against GQ$^\star$ and \emph{GQ$^\star$-routing\/}.

\subsection{Fault-tolerance}
\label{sec:setup-fault}
High reliability is of the utmost importance in datacenters, as it impacts upon the business volume that can be attracted and sustained.  When scaling out to tens of thousands of servers or more, failures are common, with the mean-time between failures (MTBF) being as short as hours or even minutes. As an example, consider a datacenter with $25,000$ servers, $1,000$ switches, and $75,000$ links, each with an optimistic average lifespan of $5$ years. Based upon a very rough estimate that the number of elements divided by the average lifespan results in the numbers of failures per day, the system will have an average of about $13$ server faults per day, $40$ link faults per day, and $1$ switch fault every $2$ days. In other words, failures are ubiquitous and so the DCN should be able to deal with them in order to remain competitively operational. Any network whose performance degrades rapidly with the number of failures is unacceptable, even if
it does provide the best performance in a fault-free environment.

We investigate how network-level failures affect \emph{routed connectivity}, defined as the proportion of server-node-pairs that remain connected by a path computable by a given routing algorithm, as well as how they affect routed mean hop-distance.  Our study focuses on uniform random link failures in particular, because both server-node and switch-node failures induce link failures, and also because the sheer number of links (and NICs) in a DCN implies that link-failures will be the most common event.  A more detailed study of failure events will be conducted in follow-up research, in which we will consider correlated link, server-node, and switch-node failures.  We consider failure configurations with up to a 15\% network degradation, where we randomly select, with uniform probability, 15\% of the links to have a fault.  Furthermore, we consider only bidirectional failures, \emph{i.e.\/}, where links will either work in both directions or in neither.  The rationale for this is that the bidirectional link-failure model is more realistic than the unidirectional one: failures affecting the whole of a link (\emph{e.g.\/}, NIC failure, unplugged or cut link, or switch-port failure) are more frequent than the fine-grained failures that would affect a single direction. In addition, once unidirectional faults have been detected they will typically be dealt with by disabling the other direction of the failed link (according to the IEEE 802.3ah EFM-OAM standard). 

As regards routed connectivity and routed mean hop-distance, we consider GQ$^\star$ with \emph{GQ$^\star$-routing\/}, FiConn with breadth-first search (BFS), and DPillar with \emph{DPillarMP\/}. Again, we explain our choice of routing algorithms. As regards FiConn, \emph{TAR\/} is a distributed ``heuristic'' algorithm devised so as to improve network load balancing with bursty and irregular traffic patterns, and was neither optimised for nor tested on outright faulty links.  In addition, \emph{TAR\/}
computes paths that are 15--30\% longer in these scenarios than \emph{TOR\/} does. However, \emph{TOR\/} is not fault-tolerant and so we simply use BFS. In short, we have given FiConn preferential treatment (this makes the performance of GQ$^\star$ against FiConn, described in Section~\ref{sec:evaluate-fault-tolerance}, all the more impressive). As regards DPillar, \emph{DPillarMP\/} is fault-tolerant whereas \emph{DPillarSP\/} is not.

\subsection{Traffic patterns}
\label{sec:traffic-patterns}

We now describe the traffic patterns used in our evaluation, the
primary one being the all-to-all traffic pattern. All-to-all communications are extremely relevant as they are intrinsic to MapReduce, the preferred paradigm for data-oriented application
development; see, for example,
\cite{LiuMuppalaVeeraraghavan2013,DeanGhemawat2008,White2009}.  In
addition, all-to-all can be considered a worst-case traffic pattern
for two reasons: ({\emph{a\/}) the lack of spatial locality; and
(\emph{b\/}) the high levels of contention for the use of resources.

Our second set of experiments focuses on specific networks hosting
around 25,000 server-nodes and evaluates them with a wider collection of traffic patterns; we use the routing algorithms \emph{GQ$^\star$-routing\/}, \emph{TOR\/}, and \emph{DPillarSP\/}.  Apart from all-to-all, we also consider the three other traffic patterns many all-to-all, butterfly, and random.  In \emph{many all-to-all\/}, the network is split into disjoint
groups of a fixed number of server-nodes with server-nodes within a group
performing an all-to-all operation.  Our evaluation shows results for
groups of 1,000 server-nodes but these results are consistent with ones for groups
of sizes 500 and 5,000.  This workload is less demanding than the
system-wide all-to-all, but can still generate a great deal of
congestion.  It aims to emulate a typical tenanted cloud datacenter
in which there are many independent applications running concurrently.  We assume
a typical topology-agnostic scheduler and randomly assign server-nodes to
groups.  The \emph{butterfly} traffic pattern is a ``logarithmic
implementation'' of a pattern such as all-to-all as each server-node only communicates with other server-nodes
at hop-distance $2^k$, for each $k\in \left\{0, \ldots,
  \left\lceil{log(N)}\right\rceil - 1 \right\}$ (see
\cite{NavaridasMiguel-AlonsoRidruejo2008} for more details).  This
workload significantly reduces the overall utilization of the network
when compared with the all-to-all traffic pattern and aims to
evaluate the behaviour of networks when the traffic pattern is
well-structured.  Finally, we consider a \emph{random} traffic pattern
in which we generate one million flows (we also studied other numbers
of flows but the results turn out to be very similar to those with one million
flows).  For each flow, the source and destination are
selected uniformly at random.  These additional
collections of experiments provide further insights into the
performance achievable with each of the networks and allow a more
detailed evaluation of propensity to congestion, load balancing, and latency.

\subsection{Software tools}
\label{sec:setuptools}

Our software tool, Interconnection Networks Research Flow Evaluation Framework (INRFlow) \cite{TEF} is specifically designed for testing large-scale, flow-based systems such as DCNs with tens or hundreds of thousands of nodes, which would prohibit the use of of packet-level simulations.  The results obtained from INRFlow inform a detailed evaluation within the intended scope of our paper.

INRFlow is capable of evaluating network topologies in two
ways. Within INRFlow we can undertake a BFS for each server-node; this allows us to compute the hop-length of the \emph{shortest path} between any two server-nodes and also to examine whether two
server-nodes become disconnected in the presence of link failures.  As we have noted in Section~\ref{sec:distance}, results on shortest paths are of limited use when not studied in conjunction with a routing algorithm. Consequently, INRFlow also provides path and connectivity information about a given routing algorithm. We use the different routing algorithms within our DCNs as we have described so far in this section. The operation of the tool is
as follows: for each flow in the workload, it computes the route using the required routing algorithm and updates link utilization
accordingly.  Then it reports a large number of statistics of
interest, including the metrics discussed above. 

\paragraph{Simulation.} Simulation is the accepted methodology as regards the empirical investigation of DCNs. For example, as regards the DCNs FiConn, HCN, BCN, SWKautz, SWCube, and SWdBruijn, all empirical analysis is undertaken by simulation; on the other hand, DCell uses a test-bed of only 20 servers, BCube uses a test-bed of only 16 servers, and CamCube \cite{Abu-LibdehCostaRowstron2010} uses a test-bed of only 27 servers. We argue that for the scenarios for which server-centric DCNs are intended, where the DCN will be expected to have thousands (if not millions) of servers (in future), experiments with a small test-bed cluster will not be
too useful (except to establish proof-of-concept) and that simulation is the best way to proceed. Moreover, the uniformity and structured design of server-centric DCNs ameliorates against performance discrepancies that might arise in ``more random'' networks.

\paragraph{Error bars.} The absence of error bars in our evaluation is by design.  In our paper, random sampling occurs in two different ways: the first is where a random set of faulty links is chosen and properties of the faulty topology are plotted, as in \Cref{plot:swnet-dpillar-routed-connectivity,plot:swnet-ficonn-unrouted-connectivity,plot:bfs_vs_routing_P_0.1,plot:bfs_vs_routing_connectivity_P_0.1};
the second is with regards to randomised traffic patterns, as in
\Cref{plot:patterns_flows,plot:patterns_dbar,plot:rnd_flows}.  For
each set of randomised link failures we plot statistics, either on
connectivity or path length, for the all-to-all traffic pattern
(\emph{i.e.,} the whole population of server-node-pairs).

In
\Cref{plot:swnet-dpillar-routed-connectivity,plot:swnet-ficonn-unrouted-connectivity,plot:bfs_vs_routing_P_0.1,plot:bfs_vs_routing_connectivity_P_0.1}
we sample the mean of two statistics over the set of all possible sets of $m$ randomised link failures based on only one trial for each network and statistic, and therefore it does not make sense to compute estimated standard error for these plots.  The true error clearly remains very small, however, because of the high level of uniformity of the DCNs we are studying, including the non-homogeneous DCN FiConn. The uniformity effectively simulates a large number of trials, since, for each choice of faulty link there are hundreds or thousands of other links in the DCN whose failure would have almost exactly the same effect on the overall experiment.  Quantifying this error is outside the scope of our paper; however, it is evident from the low amount of noise in our plots that the true error is negligible in the context of the conclusions we are making. \Cref{plot:patterns_flows,plot:patterns_dbar,plot:rnd_flows} sample flows to find the mean number of links with a certain proportion of utilisation, and to find the mean hop-lengths of the flows.  Our sample sizes, given in Section~\ref{sec:traffic-patterns}, are exceedingly large for this purpose, and thus error bars would be all but invisible in these plots.  We leave the explicit calculation to the reader.

\section{Evaluation}
\label{sec:evaluation}

In this section we perform an empirical evaluation of the DCN
GQ$^\star$ and compare its performance with that of the DCNs FiConn and DPillar using the methodology and framework as detailed in Section~\ref{sec:setup}. We begin by comparing various different versions of the three DCNs as regards ABT and latency (though the latter is a coarse-grained analysis). Next, we focus on $4$ comparable large-scale DCNs, namely GQ$^\star_{3,10}$, GQ$^\star_{4,6}$, FiConn$_{2,24}$, and DPillar$_{4,18}$, and we examine them in more detail with regard to fault-tolerance, latency, and load balancing, under different traffic patterns. Interspersed is an examination of the fault-tolerance capabilities of \emph{GQ$^\star$-routing\/} in comparison with what might happen in the optimal scenario. 

\subsection{Aggregate bottleneck throughput}
\label{sec:scalability}
We begin by comparing GQ$^\star$, FiConn, and DPillar as regards aggregate bottleneck throughput, following our framework as outlined in Section~\ref{sec:abt}; in particular, we use the routing algorithms \emph{GQ$^\star$-routing\/}, \emph{TOR\/}, and \emph{DPillarSP\/}. We work with $3$ different parameterized versions of GQ$^\star$, $2$ of FiConn, and $3$ of DPillar. Not only do we look at the relative ABT of different DCNs but we look at the scalability of each DCN in terms of ABT as the number of servers or component cost grows.

% \begin{figure}[ht]
% { \centering
% \input{cost_vs_ideal_ABT_ratio_GQSRouting_vs_TOR_DPillarSP} }
% \caption{The ratio of ABT to ideal ABT for GQ$^\star$/\emph{GQ$^\star$-routing\/}
%   DPillar/DPillarsp.}
% \label{plot:ideal-ABT-ratio}
% \end{figure}

We first consider ABT vs. the number of servers in each network.
Fig.~\ref{plot:servers_vs_ABT} shows that ABT scales much better in
GQ$^\star$~than in FiConn.  For the largest systems considered, GQ$^\star$
supports up to around three times the ABT of FiConn$_{3,n}$.  The
difference between the $3$ versions of GQ$^\star$~and FiConn$_{2,n}$ is not as large but is
still substantial. We can see that although the DCNs GQ$^\star$ are
constructed using far fewer switch-nodes and links than DPillar (when the two DCNs have the same number of server-nodes), their
maximum sustainable ABT is broadly better; indeed, the DCNs GQ$^\star_{k,n}$ with $k=2$ and $k=3$ consistently outperform all
DPillar DCNs. 

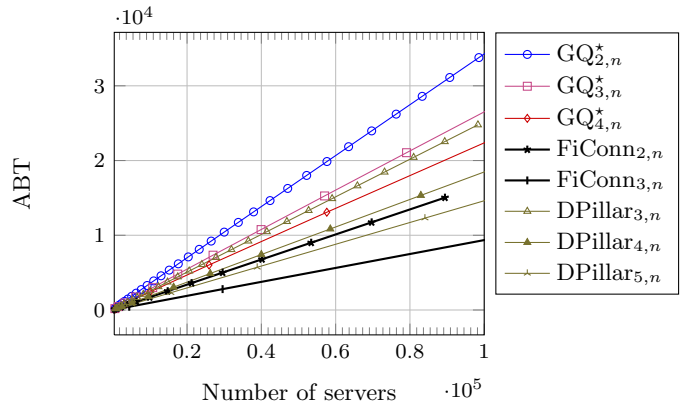
\begin{figure}[ht]
{ \centering
%
% http://sourceforge.net/projects/pgfplots/
%\usepackage{pgfplots}
%\usetikzlibrary{patterns}
%\begin{document}
% Preamble: \pgfplotsset{width=7cm,compat=1.10} %\documentclass{article} %\usepackage{pgfplots} %\pgfplotsset{compat=1.5}
% \begin{document}

%\newcommand\colourone{cyan!70!black}
%\newcommand\colourtwo{}

\begin{tikzpicture}
  \begin{axis}[width=0.5\linewidth,
    small,
    sharp plot,
   xlabel={Number of servers}, %
   ylabel={ABT},%
%   xmode=log,
   xmin=400,
   xmax=100000,
   minor x tick num = 10,
   legend entries={
     {GQ$^\star_{2,n}$},
     {GQ$^\star_{3,n}$},
     {GQ$^\star_{4,n}$},
%     {GQ$^\star_{5,n}$},
%     {GQ$^\star_{6,n}$},          
     {FiConn$_{2,n}$},
     {FiConn$_{3,n}$},
     {DPillar$_{3,n}$},
     {DPillar$_{4,n}$},
     {DPillar$_{5,n}$},
   },
   legend style={cells={anchor=west},font=\small,},
   legend pos=outer north east,
   grid=major,
   thin,
   mark size=1.5pt,
   ]

\addplot[blue,mark=o] 
        table [x=servers,  y expr=\thisrow{servers}*(\thisrow{servers}-1)/\thisrow{maxflow}]%
        {dat/all_knkstar_data_K2_P0.0.dat};
        
   \addplot [magenta!80!black,mark=square]
        table [x=servers,  y expr=\thisrow{servers}*(\thisrow{servers}-1)/\thisrow{maxflow}]%
        {dat/all_knkstar_data_K3_P0.0.dat};

   \addplot [red!80!black,mark=diamond]
   table [x=servers,  y expr=\thisrow{servers}*(\thisrow{servers}-1)/\thisrow{maxflow}]%
   {dat/all_knkstar_data_K4_P0.0.dat};

% \addplot [cyan!70!black,mark=triangle]
%         table [x=servers,  y expr=\thisrow{servers}*(\thisrow{servers}-1)/\thisrow{maxflow}]%
%         {dat/all_knkstar_data_K5_P0.0.dat};

% \addplot[brown,mark=asterisk,  ]
%         table [x=servers,  y expr=\thisrow{servers}*(\thisrow{servers}-1)/\thisrow{maxflow}]%
%         {dat/all_knkstar_data_K6_P0.0.dat};

   \addplot[black,mark=star,   thick,]
   table [x expr =\thisrow{N_sr}*1.2, y=ABT]%
   {dat/ficonn_servers_vs_ABT_K2.dat};
   \addplot[black,mark=|,thick,]
   table [x expr =\thisrow{N_sr}*1.2, y=ABT]%
   {dat/ficonn_servers_vs_ABT_K3.dat};

%old data (correct)
% \addplot[black,mark=triangle] 
%         table [x=N_sr, y=ABT]%
%         {dat/dpillar/ABT.dat};
\addplot[olive!70!black,mark=triangle]
table [x=n.servers,y= abt.N.x.N-1.over.max.flow]
{dat/dpillar_sp_alltoall_mira-sort-k-n3.dat};

\addplot[olive!70!black,mark=triangle*]
table [x=n.servers,y= abt.N.x.N-1.over.max.flow]
{dat/dpillar_sp_alltoall_mira-sort-k-n4.dat};

\addplot[olive!70!black,mark=Mercedes star]
table [x=n.servers,y= abt.N.x.N-1.over.max.flow]
{dat/dpillar_sp_alltoall_mira-sort-k-n5.dat};

\end{axis}
\end{tikzpicture}

%%% Local Variables: 
%%% mode: latex
%%% TeX-master: "GenDualPort"
%%% End: 
 %
 }
\caption{ABT using \emph{GQ$^\star$-routing\/}, \tor, and \emph{DPillarSP\/}.}
\label{plot:servers_vs_ABT}
\end{figure}

\Cref{plot:cost_vs_ABT} shows a plot of ABT vs. network cost under
the most plausible assumptions discussed in
Section~\ref{sec:network-cost}, namely that the aggregated cost of components for DPillar is around 10\% more and that of FiConn is around 2\%
less than that of GQ$^\star$.  When we normalize by network cost, we can see a similar shape to \Cref{plot:servers_vs_ABT} except that FiConn has a slightly improved scaling whereas DPillar has a slightly degraded one.

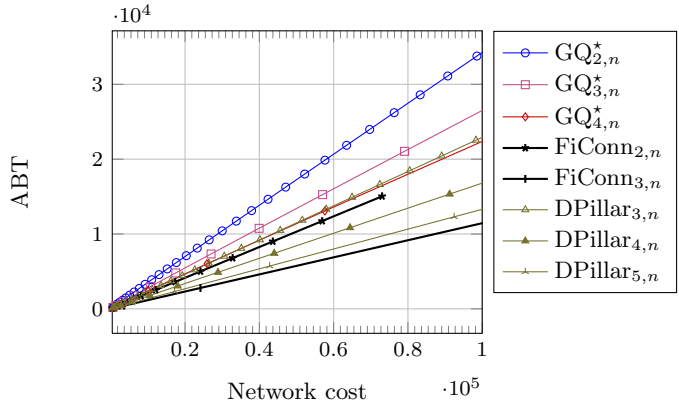
\begin{figure}[ht]
{ \centering
%
% http://sourceforge.net/projects/pgfplots/
%\usepackage{pgfplots}
%\usetikzlibrary{patterns}
%\begin{document}
% Preamble: \pgfplotsset{width=7cm,compat=1.10} %\documentclass{article} %\usepackage{pgfplots} %\pgfplotsset{compat=1.5}
% \begin{document}

%\newcommand\colourone{cyan!70!black}
%\newcommand\colourtwo{}

\begin{tikzpicture}
  \begin{axis}[width=0.5\linewidth,
    small,
    sharp plot,
   xlabel={Network cost}, %
   ylabel={ABT},%
%   xmode=log,
   xmin=400,
   xmax=100000,
   minor x tick num = 10,
   legend entries={
     {GQ$^\star_{2,n}$},
     {GQ$^\star_{3,n}$},
     {GQ$^\star_{4,n}$},
%     {GQ$^\star_{5,n}$},
%     {GQ$^\star_{6,n}$},          
     {FiConn$_{2,n}$},
     {FiConn$_{3,n}$},
     {DPillar$_{3,n}$},
     {DPillar$_{4,n}$},
     {DPillar$_{5,n}$},},
   legend style={cells={anchor=west},font=\small,},
   legend pos=outer north east,
   grid=major,
   thin,
   mark size=1.5pt,
   ]

\addplot[blue,mark=o] 
        table [x=servers,  y expr=\thisrow{servers}*(\thisrow{servers}-1)/\thisrow{maxflow}]%
        {dat/all_knkstar_data_K2_P0.0.dat};
        
   \addplot [magenta!80!black,mark=square]
        table [x=servers,  y expr=\thisrow{servers}*(\thisrow{servers}-1)/\thisrow{maxflow}]%
        {dat/all_knkstar_data_K3_P0.0.dat};

   \addplot [red!80!black,mark=diamond]
   table [x=servers,  y expr=\thisrow{servers}*(\thisrow{servers}-1)/\thisrow{maxflow}]%
   {dat/all_knkstar_data_K4_P0.0.dat};

% \addplot [cyan!70!black,mark=triangle]
%         table [x=servers,  y expr=\thisrow{servers}*(\thisrow{servers}-1)/\thisrow{maxflow}]%
%         {dat/all_knkstar_data_K5_P0.0.dat};

% \addplot[brown,mark=asterisk,  ]
%         table [x=servers,  y expr=\thisrow{servers}*(\thisrow{servers}-1)/\thisrow{maxflow}]%
%         {dat/all_knkstar_data_K6_P0.0.dat};

   \addplot[black,mark=star,   thick,]
   table [x expr =\thisrow{N_sr}*.98, y=ABT]%
    {dat/ficonn_servers_vs_ABT_K2.dat};
   \addplot[black,mark=|,thick,]
   table [x expr =\thisrow{N_sr}*.98, y=ABT]%
   {dat/ficonn_servers_vs_ABT_K3.dat};

\addplot[olive!70!black,mark=triangle]
table [x expr=1.1*\thisrow{n.servers},y= abt.N.x.N-1.over.max.flow]
{dat/dpillar_sp_alltoall_mira-sort-k-n3.dat};

\addplot[olive!70!black,mark=triangle*]
table [x expr=1.1*\thisrow{n.servers},y= abt.N.x.N-1.over.max.flow]
{dat/dpillar_sp_alltoall_mira-sort-k-n4.dat};

\addplot[olive!70!black,mark=Mercedes star]
table [x expr =1.1*\thisrow{n.servers},y= abt.N.x.N-1.over.max.flow]
{dat/dpillar_sp_alltoall_mira-sort-k-n5.dat};
% \addplot[black,mark=triangle] 
%         table [x expr =\thisrow{N_sr}*1.1, y=ABT]%
%         {dat/dpillar/ABT.dat};

\end{axis}
\end{tikzpicture}

%%% Local Variables: 
%%% mode: latex
%%% TeX-master: "GenDualPort"
%%% End: 
 %
 }
\caption{ABT in terms of network cost for \emph{GQ$^\star$-routing\/}, \tor, and
  \emph{DPillarSP\/}, where a DCN DPillar is 110\% the cost of a DCN GQ$^\star$
  with the same number of server nodes, whilst a DCN FiConn is 98\% of the
  cost of a DCN GQ$^\star$.  Network cost is normalised by the aggregated
  component cost per server in GQ$^\star$.}
\label{plot:cost_vs_ABT}
\end{figure}

Let us focus on the increase in ABT for GQ$^\star_{k,n}$ as $k$ decreases, which can be explained as follows. First, for a fixed number of server-nodes, reducing $k$ results
in an increased switch-node radix, which translates into higher locality.
Second, reducing $k$ results in lower routed mean hop-distance (see
\Cref{plot:servers_vs_hops_fixed_KP_0.0}), which lowers the total
utilization of the DCN and, when combined with good load balancing
properties, yields a bottleneck link with fewer flows. As regards routed mean hop-distances for each of the DCNs, we can see that for each topology
these
increase very slowly with network size (apart from, perhaps, FiConn$_{3,n}$) and are, of course, bounded by
the routed hop-diameter, which is dependent on $k$ for all $3$
topologies: $2k+1$ for \emph{GQ$^\star$-routing\/}; $2^{k+1}-1$ for \tor; and $2k-1$ for
\emph{DPillarSP\/}.
% not necessary to say this because we used "routed *"
% \footnote{Strictly speaking, $2^{k+1}-1$ is an upper bound on the
% diameter of FiConnpms k n.}
The ``exponential nature'' of FiConn discourages building this
topology for any $k$ larger than $2$. However, note that in terms of routed mean hop-distance, DPillar is slightly better than GQ$^\star$, broadly speaking. However, such a metric cannot be taken in isolation and we take a closer look at this metric in relation to load balancing in a more detailed evaluation of our three DCNs in Section~\ref{sec:four-dcns} (things are not what they might appear here).

\begin{figure}[ht]
{\centering
%
% http://sourceforge.net/projects/pgfplots/
%\usepackage{pgfplots}
%\usetikzlibrary{patterns}
%\begin{document}
% Preamble: \pgfplotsset{width=7cm,compat=1.10} %\documentclass{article} %\usepackage{pgfplots} %\pgfplotsset{compat=1.5}

\begin{tikzpicture}
  \begin{axis}[width=0.5\linewidth,
    small,
    xmode=log,
    sharp plot,
    xmin=100,
    xmax=100000,
   xlabel={Number of servers}, %
   ylabel={Mean hops per route},%
   legend entries={
     {GQ$^\star_{2,n}$},
%     {GQ$^\star_{2,n}$ BFS},
     {GQ$^\star_{3,n}$},
     %{GQ$^\star_{3,n}$ B},
     {GQ$^\star_{4,n}$},
     %{GQ$^\star_{4,n}$ B},
%     {GQ$^\star_{5,n}$},
     %{GQ$^\star_{5,n}$ B},
%     {GQ$^\star_{6,n}$},
%     {GQ$^\star_{6,n}$ BFS},
%     ADD FICONNS HERE
     %{FiConn$_{2,n}$ B},
     %{FiConn$_{3,n}$ B},
     {FiConn$_{2,n}$ },
     {FiConn$_{3,n}$ },
     %{DPillar$_{n,3}$ B},
     {DPillar$_{3,n}$ },
     {DPillar$_{4,n}$ },
     {DPillar$_{5,n}$ },
   },
   legend style={cells={anchor=west},font=\small,},
   legend pos=outer north east,
   grid=major,
   thin,
mark size=1.5pt,
   ]
 \addplot[blue,mark=o] table {dat/nodes_col_2_vs_hop_col_5_servers_vs_mean_hop_K2NP0.0.dat};
% \addplot [blue,mark=*] table [x expr=\thisrow{n}^\thisrow{k}*(\thisrow{n}-1)*\thisrow{k},y=meanbfs] {dat/BFS_knkstar_K2NP0.0.dat};

\addplot[magenta!80!black,mark=square] table {dat/nodes_col_2_vs_hop_col_5_servers_vs_mean_hop_K3NP0.0.dat};
% \addplot [magenta!80!black,mark=square*] table [x expr=\thisrow{n}^\thisrow{k}*(\thisrow{n}-1)*\thisrow{k},y=meanbfs] {dat/BFS_knkstar_K3NP0.0.dat};

\addplot[red!80!black,mark=diamond] table {dat/nodes_col_2_vs_hop_col_5_servers_vs_mean_hop_K4NP0.0.dat};
% \addplot [red!80!black,mark=diamond*] table [x expr=\thisrow{n}^\thisrow{k}*(\thisrow{n}-1)*\thisrow{k},y=meanbfs] {dat/BFS_knkstar_K4NP0.0.dat};

% \addplot[cyan!70!black,mark=triangle] table {dat/nodes_col_2_vs_hop_col_5_servers_vs_mean_hop_K5NP0.0.dat};
 % \addplot [cyan!70!black,mark=triangle*] table [x expr=\thisrow{n}^\thisrow{k}*(\thisrow{n}-1)*\thisrow{k},y=meanbfs] {dat/BFS_knkstar_K5NP0.0.dat};

 % \addplot[brown,mark=otimes] table {dat/nodes_col_2_vs_hop_col_5_servers_vs_mean_hop_K6NP0.1.dat};
 % \addplot [brown,mark=otimes*] table [x expr=\thisrow{n}^\thisrow{k}*(\thisrow{n}-1)*\thisrow{k},y=meanbfs] {dat/BFS_knkstar_K6NP0.1.dat};

 % FICONN TABLES GO HERE
 % NOTE THAT WE HAVE USED 0.06 IN PLACE OF 0.1.
 %THE DATA SHOULD BE RECOMPUTED AND THIS COMMENT DELETED WHEN JAVIER SENDS OUT THE NEW STUFF.
 % \addplot[red!50!black,mark=star,thick] table [x expr=\thisrow{n}^\thisrow{k}*(\thisrow{n}-1)*\thisrow{k},y=meanbfs]{dat/BFS_ficonn_K2NP0.0.dat};
 % \addplot[red!50!black,mark=|,thick] table [x expr=\thisrow{n}^\thisrow{k}*(\thisrow{n}-1)*\thisrow{k},y=meanbfs]{dat/BFS_ficonn_K3NP0.0.dat};

 \addplot[black,mark=star,thick] table []{dat/ficonn_servers_vs_mean_hop_K2NP0.0.dat};
 \addplot[black,mark=|,thick] table 
[]{dat/ficonn_servers_vs_mean_hop_K3NP0.0.dat};

\addplot[olive!70!black,mark=triangle] table
[x=n.servers,y=mean.server.hop.length]{dat/dpillar_sp_alltoall_mira-sort-k-n3.dat};

\addplot[olive!70!black,mark=triangle*] table [x=n.servers,y=mean.server.hop.length]{dat/dpillar_sp_alltoall_mira-sort-k-n4.dat};

\addplot[olive!70!black,mark=Mercedes star] table [x=n.servers,y=mean.server.hop.length]{dat/dpillar_sp_alltoall_mira-sort-k-n5.dat};

% \addplot[black,mark=triangle]
%    table [x=N_sr,y=SP]{dat/dpillar/mean_hop.dat};

 \end{axis}
\end{tikzpicture}

%%% Local Variables: 
%%% mode: latex
%%% TeX-master: "GenDualPort"
%%% End: 
 %
 }
\caption{Routed mean hop-distances for \emph{GQ$^\star$-routing\/},
  \tor, and \emph{DPillarSP\/}.}
\label{plot:servers_vs_hops_fixed_KP_0.0}
\end{figure}
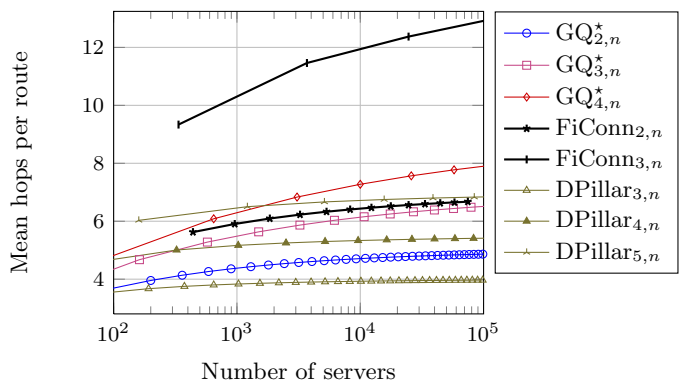

%IDEAL ABT

% DPillar/DPillarsp\ features shorter routed mean hop-distance than
% certain GQ$^\star_{3,n}$, for certain $n$, but yet it does not
% outperform it in terms of ABT, which implies that \emph{GQ$^\star$-routing\/} is
% better at load balancing than DPillarsp.  We see this plotted
% explicitly in Fig.~\ref{plot:norm_flow_hist}, but in
% Fig.~\ref{plot:ideal-ABT-ratio} we plot the ratio of ABT to ideal ABT
% vs. network cost, where the ideal ABT is given in
% Section~\ref{sec:abt}.  GQ$^\star$/\emph{GQ$^\star$-routing\/} far exceeds the
% performance of DPillar/DPillarsp/ and FiConn/\tor/ in this respect,
% and furthermore, a more careful implementation of \emph{GQ$^\star$-routing\/} could
% increase the plotted ratio considerably since it is based on
% dimensional routing, which is both optimal (distance-wise) and
% balanced in the base graph $GQ$.

Although we forgo a simulation of packet injections, our experiments do allow for a coarse-grained latency analysis.  Network latency is
brought on by routing packets over long paths and spending additional
time processing (\emph{e.g.\/}, buffering) the packets at intermediate nodes,
due to network congestion.  These scenarios have various causes, but
they are generally affected by a DCN's ability to simultaneously
balance network traffic and route it over short paths efficiently.
\Cref{plot:servers_vs_ABT,plot:servers_vs_hops_fixed_KP_0.0}
show that \emph{GQ$^\star$-routing\/} scales well with respect to load balancing
(high ABT) and routed mean hop-distance, from which we infer that in
many situations GQ$^\star_{3,n}$ has lower latency than GQ$^\star_{4,n}$ and all FiConn DCNs, and likely performs at least similarly to DPillar$_{3,n}$.

In summary, GQ$^\star$ has better ABT properties than FiConn and also broadly outperforms the denser
DPillar; as discussed in Section~\ref{sec:abt}, ABT is a performance metric of primary interest in
the context of datacenters. We can also infer from our experiments a coarse-grained latency analysis, namely that \emph{GQ$^\star$-routing\/} is likely to be at least as good as DPillar and better than FiConn.

\subsection{Fault-tolerance}
\label{sec:evaluate-fault-tolerance}

We now turn our attention to four concrete instances of the topologies and their routing algorithms: GQ$^\star_{3,10}$ and GQ$^\star_{4,6}$ with \emph{GQ$^\star$-routing\/}; FiConn$_{2,24}$ with BFS; and DPillar$_{4,18}$ with \emph{DPillarMP\/} (though we shall also consider DPillar with \emph{DPillarSP\/} in the non-fault-tolerant environment of Section~\ref{sec:four-dcns}).  As stated in Section~\ref{sec:setup-fault}, these DCNs were chosen as each has around 25,000 server-nodes and use switch-nodes with around 24 ports.  

\emph{A priori}, GQ$^\star$ has a provably high number of parallel
paths and server-parallel paths compared to FiConn and DPillar of
similar size (see Table~\ref{table:DCNs}).  Thus, if \emph{GQ$^\star$-routing\/}
utilises these paths, we expect strong performance in degraded
networks.  \Cref{plot:swnet-dpillar-routed-connectivity} shows the
routed connectivity under failures\footnote{Routed and unrouted data
  computed for other DCNs GQ$^\star$ was very similar and is not plotted
  for the sake of clarity.} of \emph{GQ$^\star$-routing\/}
and \emph{DPillarMP\/}.  The plot indicates that \emph{DPillarMP\/}
underutilises the network, since the unrouted connectivity of
DPillar (not plotted) is slightly stronger than that of GQ$^\star$.
This highlights the fact that there is a close and complex relationship between topology, path-lengths, routing, fault-tolerance, and so on; ensuring that all aspects dovetail together is of primary importance. These observations also motivate a more detailed
evaluation of \emph{GQ$^\star$-routing\/} (and indeed fault-tolerant routing for
DPillar). Note that the evaluation of \emph{DPillarMP\/} in Liao \emph{et al.}
\cite{LiaoYinYin2012} is with respect to server-node faults, in which the
performance of \emph{DPillarMP\/} looks stronger than it does in our
experiments with link-failures.  This is because the failed server-nodes do
not send messages and therefore do not factor into the connectivity of
the faulty DPillar.

\begin{figure}[ht]
  {\centering
%
% http://sourceforge.net/projects/pgfplots/
%\usepackage{pgfplots}
%\usetikzlibrary{patterns}
%\begin{document}
% Preamble: \pgfplotsset{width=7cm,compat=1.10} %\documentclass{article} %\usepackage{pgfplots} %\pgfplotsset{compat=1.5}
% \begin{document}

\begin{tikzpicture}
  \begin{axis}[width=0.5\linewidth,,
    small,
    %const plot mark mid,
    %ybar, bar width=3pt,
%    ycomb,
%    title=Histogram of number of hops,
    xmax=16,
    sharp plot,
   xlabel={Percent link failures}, %
   ylabel={Percent connectivity},%
%   x expr={\thisrow{0}*100},
   legend entries={
%     {GQ$^\star_{4,6}$},
     {GQ$^\star_{4,6}$},
     %{GQ$^\star_{3,10}$},
     {GQ$^\star_{3,10}$},
%     {FiConn$_{2,24}$},
%     {FiConn$_{3,8}$},
     {DPillar$_{4,18}$}, %MP
   },
%   legend to name=meanhops,
%   legend columns=-1,
legend style={cells={anchor=west},font=\small,},
   legend pos=outer north east,
   grid=major,
   thin,
   ]
\addplot[cyan!70!black,mark=triangle] table [x expr=\thisrow{pfailures}*100,y expr =\thisrow{pconnectivity}]  {dat/all_knkstar_data_K4_N6.dat};
%\addplot[cyan!70!black,mark=triangle*] table [x expr=\thisrow{pfailures}*100,y expr=100*\thisrow{connectedbfs}/671846400]{dat/BFS_knkstar_K4N6.dat};

\addplot[magenta!80!black,mark=o] table [x expr=\thisrow{pfailures}*100,y expr =\thisrow{pconnectivity}] {dat/all_knkstar_data_K3_N10.dat};
%\addplot[magenta!80!black,mark=*] table [x expr=\thisrow{pfailures}*100,y expr=100*\thisrow{connectedbfs}/729000000]{dat/BFS_knkstar_K3N10.dat};

%  \addplot[black,mark=star,thick]
%    table [x expr=\thisrow{pfailures}*100,y expr=100*\thisrow{connectedbfs}/607523904]{dat/BFS_ficonn_K2N24P.dat};

%  \addplot[black,mark=|,thick]
% table [x expr=\thisrow{pfailures}*100,y expr=100*\thisrow{connectedbfs}/607129600]{dat/BFS_ficonn_K3N8P.dat};

\addplot[black,mark=Mercedes star] table [x=failure_rate,y=18_4] {dat/dpillar/faulty_sp.dat};

% \addplot[black,mark=star] table [x expr=\thisrow{pfailures}*100,y expr=\thisrow{connectedbfs}/607523904]{dat/BFS_ficonn_K2N24P_PF_vs_connected.dat};
% \addplot[black,mark=|] table [x expr=\thisrow{pfailures}*100,y expr=\thisrow{connectedbfs}/607129600]{dat/BFS_ficonn_K3N8P_PF_vs_connected.dat};

\end{axis}

%\ref{meanhops}
\end{tikzpicture}
%\end{document}

%%% Local Variables: 
%%% mode: latex
%%% TeX-master: "GenDualPort"
%%% End: 
 %
}
\caption{Routed connectivity of \emph{GQ$^\star$-routing\/} and \emph{DPillarMP\/}.}
\label{plot:swnet-dpillar-routed-connectivity}
\end{figure}
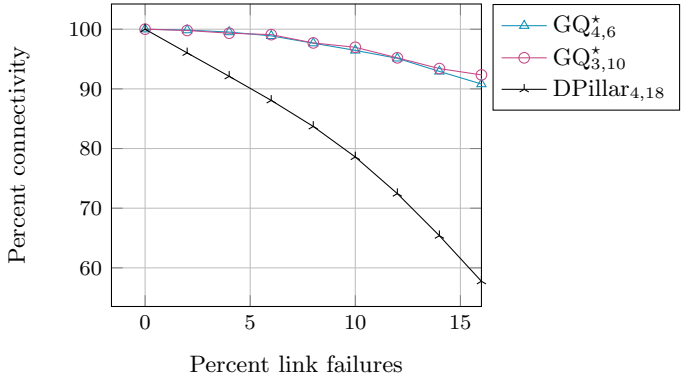

\begin{figure}[ht]
  {\centering
%
% http://sourceforge.net/projects/pgfplots/
%\usepackage{pgfplots}
%\usetikzlibrary{patterns}
%\begin{document}
% Preamble: \pgfplotsset{width=7cm,compat=1.10} %\documentclass{article} %\usepackage{pgfplots} %\pgfplotsset{compat=1.5}
% \begin{document}

\begin{tikzpicture}
  \begin{axis}[width=0.5\linewidth,,
    small,
    %const plot mark mid,
    %ybar, bar width=3pt,
%    ycomb,
%    title=Histogram of number of hops,
    xmax=16,
    sharp plot,
   xlabel={Percent link failures}, %
   ylabel={Percent connectivity},%
%   x expr={\thisrow{0}*100},
   legend entries={
%     {GQ$^\star_{4,6}$},
     {GQ$^\star_{4,6}$},
     %{GQ$^\star_{3,10}$},
     {GQ$^\star_{3,10}$},
     {FiConn$_{2,24}$},
     {FiConn$_{3,8}$},
%     {DPillar$_{18,4}$},
   },
%   legend to name=meanhops,
%   legend columns=-1,
legend style={cells={anchor=west},font=\small,},
   legend pos=outer north east,
   grid=major,
   thin,
   ]
%\addplot[cyan!70!black,mark=triangle] table [x expr=\thisrow{pfailures}*100,y expr =\thisrow{pconnectivity}]  {dat/all_knkstar_data_K4_N6.dat};
\addplot[cyan!70!black,mark=triangle*] table [x expr=\thisrow{pfailures}*100,y expr=100*\thisrow{connectedbfs}/671846400]{dat/BFS_knkstar_K4N6.dat};

%\addplot[magenta!80!black,mark=o] table [x expr=\thisrow{pfailures}*100,y expr =\thisrow{pconnectivity}] {dat/all_knkstar_data_K3_N10.dat};
\addplot[magenta!80!black,mark=*] table [x expr=\thisrow{pfailures}*100,y expr=100*\thisrow{connectedbfs}/729000000]{dat/BFS_knkstar_K3N10.dat};

 \addplot[black,mark=star,thick]
   table [x expr=\thisrow{pfailures}*100,y expr=100*\thisrow{connectedbfs}/607523904]{dat/BFS_ficonn_K2N24P.dat};

 \addplot[black,mark=|,thick]
table [x expr=\thisrow{pfailures}*100,y expr=100*\thisrow{connectedbfs}/607129600]{dat/BFS_ficonn_K3N8P.dat};

   % \addplot[black,mark=Mercedes star] table [x=failure_rate,y=18_4] {dat/dpillar/faulty_BFS.dat};

% \addplot[black,mark=star] table [x expr=\thisrow{pfailures}*100,y expr=\thisrow{connectedbfs}/607523904]{dat/BFS_ficonn_K2N24P_PF_vs_connected.dat};
% \addplot[black,mark=|] table [x expr=\thisrow{pfailures}*100,y expr=\thisrow{connectedbfs}/607129600]{dat/BFS_ficonn_K3N8P_PF_vs_connected.dat};

\end{axis}

%\ref{meanhops}
\end{tikzpicture}
%\end{document}

%%% Local Variables: 
%%% mode: latex
%%% TeX-master: "GenDualPort"
%%% End: 
 %
}
\caption{Unrouted connectivity of GQ$^\star$ and FiConn.}
\label{plot:swnet-ficonn-unrouted-connectivity}
\end{figure}
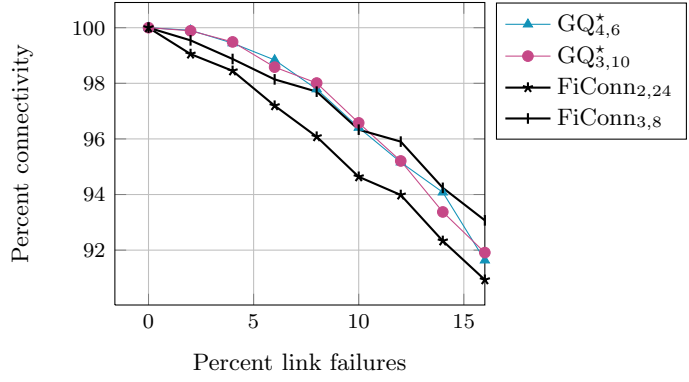

\subsection{Assessment of \emph{GQ$^\star$-routing\/}}
\label{sec:evaluate-gqsrouting}

With FiConn not having a fault-tolerant algorithm comparable to
\emph{GQ$^\star$-routing\/} (see Section~\ref{sec:setuptools}), in
\Cref{plot:swnet-ficonn-unrouted-connectivity} we plot the
unrouted connectivity of GQ$^\star$ with that of FiConn using BFS. As we can see, \emph{GQ$^\star$-routing\/} performs similarly to FiConn in an optimal scenario.  To our knowledge
there is no fault-tolerant routing algorithm for FiConn that
achieves anything close to the optimal performance of BFS (however,
\Cref{plot:bfs_vs_routing_connectivity_P_0.1} shows that
\emph{GQ$^\star$-routing\/} very nearly achieves the optimum unrouted
connectivity of GQ$^\star$).

In summary, we have shown that GQ$^\star$ and \emph{GQ$^\star$-routing\/} are very competitive when compared with both FiConn and DPillar in terms of fault-tolerance.

\begin{figure}[ht]
{\centering
%
% http://sourceforge.net/projects/pgfplots/
%\usepackage{pgfplots}
%\usetikzlibrary{patterns}
%\begin{document}
% Preamble: \pgfplotsset{width=7cm,compat=1.10} %\documentclass{article} %\usepackage{pgfplots} %\pgfplotsset{compat=1.5}

\begin{tikzpicture}
  \begin{axis}[width=0.5\linewidth,
    small,
    xmode=log,
    sharp plot,
    xmax=100000,
   xlabel={Number of servers}, %
   ylabel={Mean hops per route},%
   legend entries={
%     {GQ$^\star_{2,n}$},
%     {GQ$^\star_{2,n}$ BFS},
     {GQ$^\star_{3,n}$},
     {GQ$^\star_{3,n}$ BFS},
     {GQ$^\star_{4,n}$},
     {GQ$^\star_{4,n}$ BFS},
     {GQ$^\star_{5,n}$},
     {GQ$^\star_{5,n}$ BFS},
%     {GQ$^\star_{6,n}$},
%     {GQ$^\star_{6,n}$ BFS},
%     ADD FICONNS HERE
%     {FiConn$_{2,n}$},
%     {FiConn$_{3,n}$},
   },
   legend style={cells={anchor=west},font=\small,},
   legend pos=outer north east,
   grid=major,
   thin,
mark size=1.5pt,
   ]
% \addplot[blue,mark=o] table {dat/nodes_col_2_vs_hop_col_5_servers_vs_mean_hop_K2NP0.1.dat};
% \addplot [blue,mark=*] table [x expr=\thisrow{n}^\thisrow{k}*(\thisrow{n}-1)*\thisrow{k},y=meanbfs] {dat/BFS_knkstar_K2NP0.1.dat};

   \addplot[magenta!80!black,mark=square]
   table {dat/nodes_col_2_vs_hop_col_5_servers_vs_mean_hop_K3NP0.1.dat};
   \addplot [magenta!80!black,mark=square*]
   table [x expr=\thisrow{n}^\thisrow{k}*(\thisrow{n}-1)*\thisrow{k},y=meanbfs] {dat/BFS_knkstar_K3NP0.1.dat};

   \addplot[red!80!black,mark=diamond]
   table {dat/nodes_col_2_vs_hop_col_5_servers_vs_mean_hop_K4NP0.1.dat};
   \addplot [red!80!black,mark=diamond*]
   table [x expr=\thisrow{n}^\thisrow{k}*(\thisrow{n}-1)*\thisrow{k},y=meanbfs] {dat/BFS_knkstar_K4NP0.1.dat};

   \addplot[cyan!70!black,mark=triangle]
   table {dat/nodes_col_2_vs_hop_col_5_servers_vs_mean_hop_K5NP0.1.dat};
   \addplot [cyan!70!black,mark=triangle*]
   table [x expr=\thisrow{n}^\thisrow{k}*(\thisrow{n}-1)*\thisrow{k},y=meanbfs] {dat/BFS_knkstar_K5NP0.1.dat};

 % \addplot[brown,mark=otimes] table {dat/nodes_col_2_vs_hop_col_5_servers_vs_mean_hop_K6NP0.1.dat};
 % \addplot [brown,mark=otimes*] table [x expr=\thisrow{n}^\thisrow{k}*(\thisrow{n}-1)*\thisrow{k},y=meanbfs] {dat/BFS_knkstar_K6NP0.1.dat};

 % FICONN TABLES GO HERE
 % NOTE THAT WE HAVE USED 0.06 IN PLACE OF 0.1.
 %THE DATA SHOULD BE RECOMPUTED AND THIS COMMENT DELETED WHEN JAVIER SENDS OUT THE NEW STUFF.
   % \addplot[black,mark=star]
   % table [x expr=\thisrow{n}^\thisrow{k}*(\thisrow{n}-1)*\thisrow{k},y=meanbfs]{dat/BFS_ficonn_K2NP0.1.dat};
   % \addplot[black,mark=|]
   % table [x expr=\thisrow{n}^\thisrow{k}*(\thisrow{n}-1)*\thisrow{k},y=meanbfs]{dat/BFS_ficonn_K3NP0.1.dat};

\end{axis}
\end{tikzpicture}

%%% Local Variables: 
%%% mode: latex
%%% TeX-master: "GenDualPort"
%%% End: 
 %
}
\caption{Routed (\emph{GQ$^\star$-routing\/}) and unrouted mean-distance in GQ$^\star$
  with $10\%$ link failures.}
\label{plot:bfs_vs_routing_P_0.1}
\end{figure}
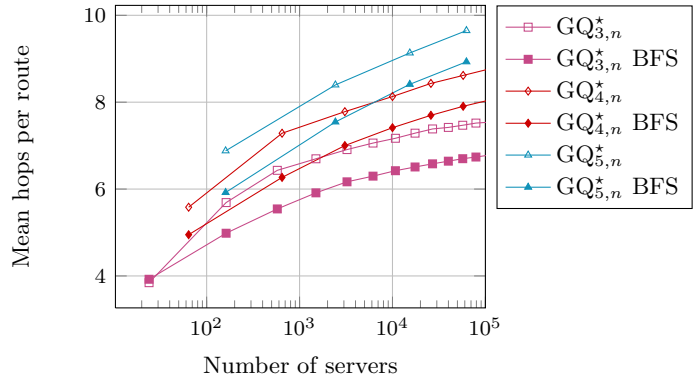

We assess the performance of \emph{GQ$^\star$-routing\/} by comparing it with
optimum performance, obtained by computing a BFS which finds a shortest-path (if
it exists).
Notice that since dimensional routing yields a shortest-path algorithm on
$GQ_{k,n}$, it is straightforward to modify \emph{GQ$^\star$-routing\/} so as to
be a shortest path algorithm on GQ$^\star_{k,n}$; however, due to
simplifications in our implementation there is a discrepancy of about
2\% between shortest paths and \emph{GQ$^\star$-routing\/} in a fault-free GQ$^\star$.

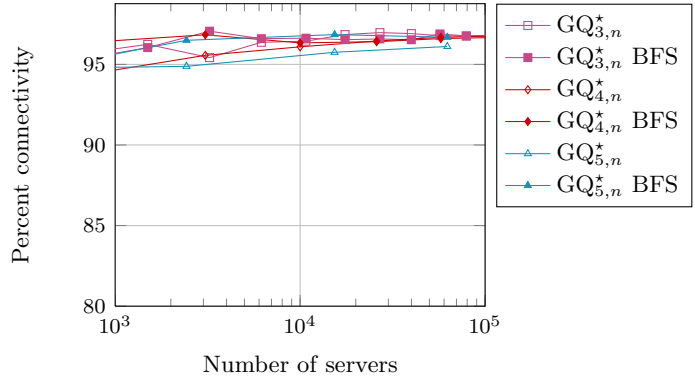
\begin{figure}[ht]
{\centering
  %
% http://sourceforge.net/projects/pgfplots/
%\usepackage{pgfplots}
%\usetikzlibrary{patterns}
%\begin{document}
% Preamble: \pgfplotsset{width=7cm,compat=1.10} %\documentclass{article} %\usepackage{pgfplots} %\pgfplotsset{compat=1.5}

\begin{tikzpicture}
  \begin{axis}[width=0.5\linewidth,
    small,
    xmode=log,
    sharp plot,
    xmin=1000,
    xmax=100000,
ymin=80,
   xlabel={Number of servers}, %
   ylabel={Percent connectivity},%
   legend entries={
%     {GQ$^\star_{2,n}$},
%     {GQ$^\star_{2,n}$ BFS},
     {GQ$^\star_{3,n}$},
     {GQ$^\star_{3,n}$ BFS},
     {GQ$^\star_{4,n}$},
     {GQ$^\star_{4,n}$ BFS},
     {GQ$^\star_{5,n}$},
     {GQ$^\star_{5,n}$ BFS},
%    {GQ$^\star_{6,n}$},
%    {GQ$^\star_{6,n}$ BFS},
    % ADD FICONNS HERE
    % {FiConn\ $_{2,n}$},
    % {FiConn$_{3,n}$},
  },
   legend style={cells={anchor=west},font=\small,},
   legend pos=outer north east,
   grid=major,
   thin,
mark size=1.5pt,
   ]
% \addplot[blue,mark=o] table {dat/nodes_col_2_vs_hop_col_5_servers_vs_mean_hop_K2NP0.1.dat};
% \addplot [blue,mark=*] table [x expr=\thisrow{n}^\thisrow{k}*(\thisrow{n}-1)*\thisrow{k},y=meanbfs] {dat/BFS_knkstar_K2NP0.1.dat};

   \addplot [magenta!80!black,mark=square]
   table [x=servers,y expr=\thisrow{pconnectivity}]
   {dat/routing_knkstar_K3NP0.1.dat};

   \addplot [magenta!80!black,mark=square*]
   table [x expr=\thisrow{n}^\thisrow{k}*(\thisrow{n}-1)*\thisrow{k},y expr=(\thisrow{connectedbfs}
   /((\thisrow{n}^\thisrow{k})*(\thisrow{n}-1)*\thisrow{k})^2)*100]
   {dat/BFS_knkstar_K3NP0.1.dat};

   \addplot[red!80!black,mark=diamond]
   table [x=servers,y expr=\thisrow{pconnectivity}]
   {dat/routing_knkstar_K4NP0.1.dat};

   \addplot [red!80!black,mark=diamond*]
   table [x expr=\thisrow{n}^\thisrow{k}*(\thisrow{n}-1)*\thisrow{k},y expr=(\thisrow{connectedbfs}
   /((\thisrow{n}^\thisrow{k})*(\thisrow{n}-1)*\thisrow{k})^2)*100]
   {dat/BFS_knkstar_K4NP0.1.dat};

   \addplot[cyan!70!black,mark=triangle]
   table [x=servers,y expr=\thisrow{pconnectivity}]
   {dat/routing_knkstar_K5NP0.1.dat};

   \addplot [cyan!70!black,mark=triangle*]
   table [x expr=\thisrow{n}^\thisrow{k}*(\thisrow{n}-1)*\thisrow{k},y expr=(\thisrow{connectedbfs}
   /((\thisrow{n}^\thisrow{k})*(\thisrow{n}-1)*\thisrow{k})^2)*100]
   {dat/BFS_knkstar_K5NP0.1.dat};

   %       \addplot [green!80!black,mark=o]
   % table [x=servers,y expr=\thisrow{pconnectivity}]
   % {dat/routing_knkstar_K6NP0.1.dat};

   % \addplot [green!80!black,mark=*]
   % table [x expr=\thisrow{n}^\thisrow{k}*(\thisrow{n}-1)*\thisrow{k},y expr=(\thisrow{connectedbfs}
   % /((\thisrow{n}^\thisrow{k})*(\thisrow{n}-1)*\thisrow{k})^2)*100]
   % {dat/BFS_knkstar_K6NP0.1.dat};

   % \addplot[red!80!black,mark=diamond]
   % table {dat/nodes_col_2_vs_hop_col_5_servers_vs_mean_hop_K4NP0.1.dat};
   % \addplot [red!80!black,mark=diamond*]
   % table [x expr=\thisrow{n}^\thisrow{k}*(\thisrow{n}-1)*\thisrow{k},y=meanbfs] {dat/BFS_knkstar_K4NP0.1.dat};

   % \addplot[cyan!70!black,mark=triangle]
   % table {dat/nodes_col_2_vs_hop_col_5_servers_vs_mean_hop_K5NP0.1.dat};
   % \addplot [cyan!70!black,mark=triangle*]
   % table [x expr=\thisrow{n}^\thisrow{k}*(\thisrow{n}-1)*\thisrow{k},y=meanbfs] {dat/BFS_knkstar_K5NP0.1.dat};

 % \addplot[brown,mark=otimes] table {dat/nodes_col_2_vs_hop_col_5_servers_vs_mean_hop_K6NP0.1.dat};
 % \addplot [brown,mark=otimes*] table [x expr=\thisrow{n}^\thisrow{k}*(\thisrow{n}-1)*\thisrow{k},y=meanbfs] {dat/BFS_knkstar_K6NP0.1.dat};

 % FICONN TABLES GO HERE
 % NOTE THAT WE HAVE USED 0.06 IN PLACE OF 0.1.
 %THE DATA SHOULD BE RECOMPUTED AND THIS COMMENT DELETED WHEN JAVIER SENDS OUT THE NEW STUFF.
   % \addplot[black,mark=star]
   % table [x expr=\thisrow{n}^\thisrow{k}*(\thisrow{n}-1)*\thisrow{k},y=meanbfs]{dat/BFS_ficonn_K2NP0.1.dat};
   % \addplot[black,mark=|]
   % table [x expr=\thisrow{n}^\thisrow{k}*(\thisrow{n}-1)*\thisrow{k},y=meanbfs]{dat/BFS_ficonn_K3NP0.1.dat};

\end{axis}
\end{tikzpicture}

%%% Local Variables: 
%%% mode: latex
%%% TeX-master: "GenDualPort"
%%% End: 
 %
}
\caption{Routed (\emph{GQ$^\star$-routing\/}) and unrouted connectivity of GQ$^\star$
  with $10\%$ link failures.}
\label{plot:bfs_vs_routing_connectivity_P_0.1}
\end{figure}

Of interest to us here is the relative performance of \emph{GQ$^\star$-routing\/}
and BFS in faulty networks.  \Cref{plot:bfs_vs_routing_P_0.1}
plots the routed and unrouted mean hop-distances in networks with a
$10\%$ link failure rate; as can be seen, the difference between \emph{GQ$^\star$-routing\/}
and BFS in mean hop-distance is close to $10\%$.  This is a reasonable
overhead for a fault-tolerant routing algorithm, especially given the algorithm's
high success rate at connecting pairs of servers in faulty networks: 
\Cref{plot:bfs_vs_routing_connectivity_P_0.1} plots the unrouted
connectivity, which is optimum and achieved by a BFS, and the routed
connectivity, achieved by \emph{GQ$^\star$-routing\/}, for the same (10\%) failure
rate\footnote{\emph{GQ$^\star$-routing\/}~appears to be better than BFS for certain
  numbers of servers, but this is because the faults were generated
  randomly for each test.}.  As it is currently implemented, \emph{GQ$^\star$-routing\/} is
optimised for maintaining connectivity at the cost of routing over
longer paths if necessary.  A different mix of features might reduce the
$10\%$ gap in \Cref{plot:bfs_vs_routing_P_0.1} but increase the
gap in \Cref{plot:bfs_vs_routing_connectivity_P_0.1}.  In any
case, the performance of \emph{GQ$^\star$-routing\/} is very close to the optimum.

%[REDUNDANT]
% \emph{GQ$^\star$-routing\/} is (potentially) optimum for fault-free networks, and
% in addition it is able to support the communication of most of the
% connected pairs of servers when failures are considered, with only a
% $10\%$ overhead in average distance between pairs of servers, which is
% a low price to pay for such resilient routing.

% \ifdraft
%[JAVIER: I wrote near-optimal for consistency, as we show there is a 2% divergence between BFS and \emph{GQ$^\star$-routing\/}]
%\fi

%THIS WAS DUPLICATED
% \begin{figure}[!ht]
%   \input{servers_vs_hops_fixed_KP_0.0}
% \caption{Average distance of shortest path and \emph{GQ$^\star$-routing\/} -- no failures.}
% \label{plot:bfs_vs_routing_P_0.0}
% \end{figure}

\subsection{Detailed evaluation of large-scale DCNs}
\label{sec:four-dcns}

We now return to our four concrete instances of the topologies and
their basic routing algorithms: GQ$^\star_{3,10}$ and GQ$^\star_{4,6}$ with
\emph{GQ$^\star$-routing\/}; FiConn$_{2,24}$ with \tor; and DPillar$_{4,18}$ with
\emph{DPillarSP\/}. Our intention is to look at throughput, how loads are balanced, and the impact on latency.

\Cref{plot:patterns_flows} shows the number of flows in the 
bottleneck for the different traffic
patterns considered in our study.  We can see that these results follow those described above in that not only can GQ$^\star$ broadly outperform FiConn and DPillar in terms of ABT, cost, latency, and fault-tolerance, but it does likewise in terms of throughput in that it can significantly reduce the
number of flows in the bottleneck.
The only exception is DPillar$_{4,18}$
with the butterfly traffic pattern.  The rationale for these results is that
the butterfly pattern matches perfectly the DPillar topology and,
thus, it allows a very good balancing of the network, reducing the
flows in the bottleneck.  For the rest of the patterns, DPillar is
clearly the worst performing in terms of bottleneck flow.
\Cref{plot:patterns_dbar} shows the routed mean hop-distance for the
different patterns and topologies, and shows that DPillar, due to the
higher number of switches, can generally reach its destination using
the shortest paths.  Note that even with the clear advantage of having higher
availability of shorter paths, DPillar$_{4,18}$ still has the highest
number of flows in the bottleneck and, therefore, is the most prone to
congestion.  On the other hand GQ$^\star_{4,6}$, which uses the longest
paths, has the second lowest number of flows in the bottleneck after
GQ$^\star_{3,10}$.
%Add latency discussion here once the new data is ready.

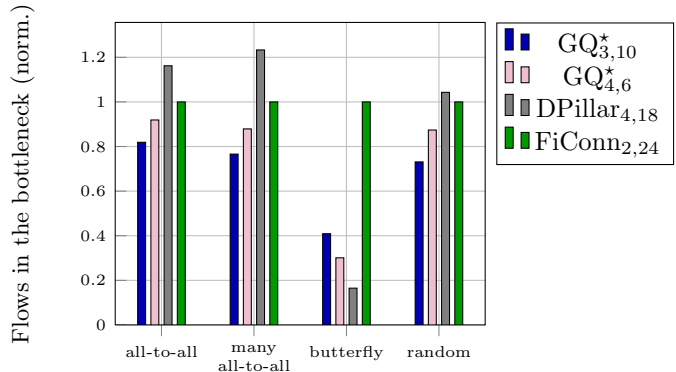
\begin{figure}[ht]
  \centering
%
% http://sourceforge.net/projects/pgfplots/
%\usepackage{pgfplots}
%\usetikzlibrary{patterns}
%\begin{document}
% Preamble: \pgfplotsset{width=7cm,compat=1.10} %\documentclass{article} %\usepackage{pgfplots} %\pgfplotsset{compat=1.5}

\begin{tikzpicture}
  \begin{axis}[width=0.5\linewidth,
    small,
	ybar,
    tick label style={font=\tiny},
    tickpos=left,
		xticklabel style={align=center},
    xticklabels={all-to-all, {many\\all-to-all}, butterfly, random}, 
    xtick={1,2,3,4},
		ylabel={Flows in the bottleneck (norm.)},
    ymin=0,
		xmin=0.5,
		xmax=4.5,
		bar width=3,
		grid=major,
    legend entries={
%     {GQ$^\star_{2,n}$},
%     {GQ$^\star_{2,n}$ BFS},
     {GQ$^\star_{3,10}$},
     {GQ$^\star_{4,6}$},
     {DPillar$_{4,18}$},
     {FiConn$_{2,24}$},
%     {FiConn$_{3,n}$},
   },
    y tick label style={/pgf/number format/.cd,%
          scaled y ticks = false,
          set thousands separator={},
          fixed
    },
		legend pos=outer north east,
    ]
    \addplot[fill=\knkthreecol]
    table {dat/flows_patterns_GQS_3_10.dat};
   
	 \addplot[fill=\knkfourcol]
    table {dat/flows_patterns_GQS_4_6.dat};

   \addplot[fill=gray]
    table {dat/flows_patterns_dpillar_18_4.dat};

   \addplot[fill=\ficonncol]
		table {dat/flows_patterns_ficonn_2_24.dat};
\end{axis}
\end{tikzpicture}

 \caption{Relative number of flows in the bottleneck for the different traffic patterns, normalised to FiConn and \emph{TOR\/}.
  %GQ$^\star_{3,10}$, GQ$^\star_{4,6}$, and FiConn$_{2,24}$,
  %DPillar$_{4,18}$ SP, respectively.
}
\label{plot:patterns_flows}
\end{figure}

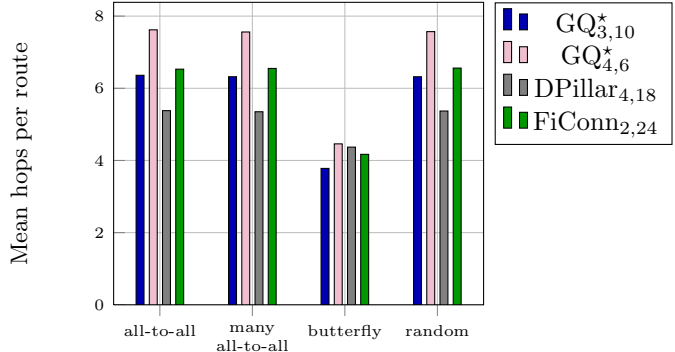
\begin{figure}[ht]
  \centering
%
% http://sourceforge.net/projects/pgfplots/
%\usepackage{pgfplots}
%\usetikzlibrary{patterns}
%\begin{document}
% Preamble: \pgfplotsset{width=7cm,compat=1.10} %\documentclass{article} %\usepackage{pgfplots} %\pgfplotsset{compat=1.5}

\begin{tikzpicture}
  \begin{axis}[width=0.5\linewidth,
    small,
	ybar ,
    tick label style={font=\tiny},
    tickpos=left,
		xticklabel style={align=center},
    xticklabels={all-to-all, {many\\all-to-all}, butterfly, random}, 
    xtick={1,2,3,4},
		ylabel={Mean hops per route},
    ymin=0,
		xmin=0.5,
		xmax=4.5,
		bar width=3,
    legend entries={
%     {GQ$^\star_{2,n}$},
%     {GQ$^\star_{2,n}$ BFS},
     {GQ$^\star_{3,10}$},
     {GQ$^\star_{4,6}$},
     {DPillar$_{4,18}$},
     {FiConn$_{2,24}$},
%     {FiConn$_{3,n}$},
   },
    y tick label style={/pgf/number format/.cd,%
          scaled y ticks = false,
          set thousands separator={},
          fixed
    },grid=major,
		legend pos=outer north east,
    ]
    \addplot[fill=\knkthreecol]
    table {dat/dbar_patterns_GQS_3_10.dat};
   
	 \addplot[fill=\knkfourcol]
    table {dat/dbar_patterns_GQS_4_6.dat};

   \addplot[fill=gray]
    table {dat/dbar_patterns_dpillar_18_4.dat};

   \addplot[fill=\ficonncol]
		table {dat/dbar_patterns_ficonn_2_24.dat};
\end{axis}
\end{tikzpicture}

 \caption{Routed hop-distance for the different traffic patterns.
  %GQ$^\star_{3,10}$, GQ$^\star_{4,6}$, and FiConn$_{2,24}$,
  %DPillar$_{4,18}$ SP, respectively.
}
\label{plot:patterns_dbar}
\end{figure}

The results we have obtained as regards bottleneck flows and routed hop-distances might appear surprising. However, a closer analysis helps to better appreciate the situation. \Cref{plot:norm_flow_hist} shows the distribution of flows across links in the all-to-all traffic pattern: for a given number of flows, we show the proportion of links carrying that number of flows. We can see that both GQ$^\star$s are much better balanced than both FiConn$_{2,24}$ and DPillar$_{4,18}$. For example, in GQ$^\star_{3,10}$ all of the links carry between $60,000$ and $100,000$ flows, and in GQ$^\star_{4,6}$ all of the links carry between $80,000$ and $120,000$ flows. However, nearly $25\%$ of the links in
FiConn$_{2,24}$ have less than $40,000$ flows, whereas the
other 75\% of the links have between $80,000$ and $140,000$ flows. Even worse, in DPillar$_{4,18}$  half of the links
have more than $100,000$ flows while the other half are barely used. The imbalances present in FiConn$_{2,24}$ and DPillar$_{4,18}$ result in parts of the networks being significantly underutilised and other parts being overly congested.

A more detailed distribution obtained using the random traffic pattern
is shown in \Cref{plot:rnd_flows}.  Here, we can see how both
GQ$^\star$s are clearly better balanced than FiConn$_{2,24}$, as the
latter has two pinnacles: one of low-load with about 30\% of the links,
and another of high-load with the rest of the links.  We can also see
that choosing the bottleneck link as the figure of merit is reasonable
as it would yield similar results as if we had chosen the peaks in the
plot.

Just as we did in Section \ref{sec:scalability}, we can infer that GQ$^\star_{3,10}$
will provide better latency figures than GQ$^\star_{4,6}$ and
FiConn$_{2,24}$ as it has fewer flows in the bottleneck link and
uses shorter paths.  The shorter paths in DPillar$_{4,18}$ do suggest that
with low-intensity communication workloads it should have lower
latency than GQ$^\star_{3,10}$, but since DPillar$_{4,18}$ is much poorer at
balancing loads than GQ$^\star_{3,10}$, we can infer that it may have
higher latency under higher-intensity communication workloads such
as the ones typically used in datacenters.

\section{Conclusion}
\label{sec:conclusion}

This paper proposes a new, generic construction that can be used to automatically convert existing interconnection networks, and their properties in relation to routing, path length, node-disjoint paths, and so on, into dual-port server-centric DCNs, that inherit the properties of the interconnection network. A range of interconnection networks has been identified to which our construction might be applied. A particular instantiation of our construction, the DCN GQ$^\star$ where the base interconnection network is the generalized hypercube, has been empirically validated as regards network throughput, latency, load balancing capability, fault-tolerance, and cost to build. In particular, we have shown how GQ$^\star$, with its routing algorithm \emph{GQ$^\star$-routing\/}, that is inherited from an existing routing algorithm for the generalized hypercube, consistently outperforms the established DCNs FiConn, with its routing algorithm \emph{TOR\/}, and DPillar, with its routing algorithms \emph{DPillarSP\/} and \emph{DPillarMP\/}. As regards FiConn, the improved performance of GQ$^\star$ was across all of the metrics we studied, apart from aggregated component cost where the two DCNs were approximately equal. As regards DPillar, the improved performance was across all metrics, apart from mean routed hop-distance. However, in mitigation against DPillar's improved mean routed hop-distance, our experiments as regards load balancing enable us to infer that although DPillar will exhibit lower latency in the case of low traffic congestion, when there is average to high traffic congestion DPillar's propensity to unbalanced loads on its links will mean that GQ$^\star$ will have the better latency. Particularly marked improvements of GQ$^\star$ against DPillar are as regards the fault-tolerant performance of the respective routing algorithms in link-degraded DCNs and also the aggregated component cost which in DPillar is around 10\% higher than in GQ$^\star$. When we compare the performance of \emph{GQ$^\star$-routing\/} within GQ$^\star$ against what is optimally possible, in terms of path length, we find that \emph{GQ$^\star$-routing\/} finds
paths that are within $2\%$ of the optimal length ($0\%$ is
realistically possible) and within around $10\%$ for degraded networks with $10\%$ faulty links.  This is a relatively small overhead for our
routing algorithm which achieves very high connectivity, typically
$95\%$ connectivity when $10\%$ of links are chosen to be faulty (uniformly at random).

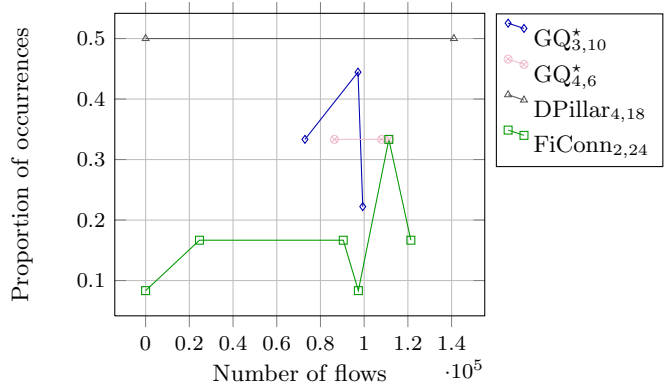
\begin{figure}[t]
  \centering
%
% http://sourceforge.net/projects/pgfplots/
%\usepackage{pgfplots}
%\usetikzlibrary{patterns}
%\begin{document}
% Preamble: \pgfplotsset{width=7cm,compat=1.10} %\documentclass{article} %\usepackage{pgfplots} %\pgfplotsset{compat=1.5}
% \begin{document}
\newcommand{\fillopacityxyz}{1}

\begin{tikzpicture}
  \begin{axis}[width=0.75\linewidth,
    % const plot mark left,
    small,
sharp plot,
    ybar, bar width=3pt,
%    ycomb,
%   title=Histogram of number of hops,
   xlabel={Number of flows},
   ylabel={Proportion of occurrences},
   legend entries={
     {GQ$^\star_{3,10}$},% $\overline x=89567$,
     {GQ$^\star_{4,6}$},% $\overline x =101953$,
     {DPillar$_{4,18}$},
     {FiConn$_{2,24}$},% $\overline x= 84615$
     %{DPillar$_{18,4}$ MP},
   },
   legend style={cells={anchor=west},font=\small,},
   legend pos=outer north east,
   grid=major,
   thin,
   mark size=1.5pt,
   ]

   \addplot[\knkthreecol,mark=diamond]
    table[x=fi,y=nf] {dat/alltoall/knkstar_interintra_k3n10r4_seed13_histograms.dat};
   
	 \addplot[\knkfourcol,mark=otimes]
    table[x=fi,y=nf] {dat/alltoall/knkstar_interintra_k4n6r4_seed13_histograms.dat};
%rnd_flows_GQS_4_6.dat};

    \addplot[gray!70!black,mark=triangle] table [x=fi,y=nf]
    {dat/alltoall/dpillar_sp_n18k4_seed13_histograms.dat};

    \addplot[\ficonncol,mark=square] table [x=fi,y=nf]
    {dat/alltoall/ficonn_tor_k2n24_seed13_histograms.dat};

   % \addplot[fill=blue] table {dat/flow_hist_K3N10P0.0S.dat};
%\addplot[draw=blue] table {dat/knkstarK3N10P0_0S13219R4_hop_histogram.dat};
%\addplot[fill=purple] table {dat/flow_hist_K4N6P0.0S.dat};
%\addplot[fill=green] table {dat/flow_hist_ficonn_K2N24.dat};

    %replacing this on feb 12, 2016
%    \addplot[fill=\knkthreecol, fill opacity=\fillopacityxyz ] table
%    {dat/norm_bucket_flow_hist_K3N10P0.0S.dat};%
%    %{dat/flow_hist_K3N10P0.0S.dat};%
% \addplot[fill=\knkfourcol, fill opacity=\fillopacityxyz ] table
%    {dat/norm_bucket_flow_hist_K4N6P0.0S.dat};%
%    %{dat/flow_hist_K4N6P0.0S.dat};%
%    \addplot[fill=gray, fill opacity=\fillopacityxyz] table
%    [x=flow,y=SP]{dat/dpillar/flow_hist.dat};%
% %{dat/dpillar/flow_hist.dat};%
%    \addplot[fill=\ficonncol, fill opacity=\fillopacityxyz] table
%    {dat/norm_bucket_flow_hist_ficonn_K2N24.dat};%
%    %{dat/flow_hist_ficonn_K2N24.dat};%

      %\addplot[fill=white, fill opacity=\fillopacityxyz] table
   %[x=flow,y=MP]{dat/dpillar/flow_hist.dat};%

   % \addplot[draw=purple] table {dat/knkstarK4N6P0.0S16331R4_hop_histogram.dat};
\end{axis}
\end{tikzpicture}
%\end{document}

%%% Local Variables: 
%%% mode: latex
%%% TeX-master: "GenDualPort"
%%% End: 
 \caption{Histogram of proportion of flows per link under
  the all-to-all traffic pattern. The mean number of flows per link
  are $89,567$, $101,953$, $84,615$, and $141,187$, for GQ$^\star_{3,10}$,
  GQ$^\star_{4,6}$, FiConn$_{2,24}$ and DPillar$_{4,18}$,
  respectively.  Connecting lines are drawn for clarity.}
\label{plot:norm_flow_hist}
\end{figure}

\begin{figure}[t]
  \centering
%
% http://sourceforge.net/projects/pgfplots/
%\usepackage{pgfplots}
%\usetikzlibrary{patterns}
%\begin{document}
% Preamble: \pgfplotsset{width=7cm,compat=1.10} %\documentclass{article} %\usepackage{pgfplots} %\pgfplotsset{compat=1.5}

\begin{tikzpicture}
  \begin{axis}[width=0.5\linewidth,
    small,
    %xmode=log,
    sharp plot,
ymax=2000,
   xlabel={Number of flows}, %
   ylabel={Number of links},%
   legend entries={
%     {GQ$^\star_{2,n}$},
%     {GQ$^\star_{2,n}$ BFS},
     {GQ$^\star_{3,10}$},
     {GQ$^\star_{4,6}$},
     {DPillar$_{4,18}$},
     {FiConn$_{2,24}$},
%     {FiConn$_{3,n}$},
   },
   legend style={cells={anchor=west},font=\small,},
   legend pos=outer north east,
   grid=major,
   thin,
   draw=none,
   mark size=1.5pt,
   ]

   \addplot[only marks,\knkthreecol,mark=diamond]
    table[x=fi,y=f] {dat/random/knkstar_interintra_k3n10r4_seed13_histograms.dat};
   
	 \addplot[only marks,\knkfourcol,mark=otimes]
    table[x=fi,y=f] {dat/random/knkstar_interintra_k4n6r4_seed13_histograms.dat};
%rnd_flows_GQS_4_6.dat};

    \addplot[only marks,gray!70!black,mark=triangle] table [x=fi,y=f]
    {dat/random/dpillar_sp_n18k4_seed13_histograms.dat};

    \addplot[only marks,\ficonncol,mark=square] table [x=fi,y=f]
    {dat/random/ficonn_tor_k2n24_seed13_histograms.dat};

       % \addplot[only marks,\knkthreecol,mark=diamond]
   %  table {dat/rnd_flows_GQS_3_10.dat};

    % 	 \addplot[only marks,\knkfourcol,mark=otimes]
    % table {dat/rnd_flows_GQS_4_6.dat};

    % \addplot[only marks,gray!70!black,mark=triangle] table
    % {dat/rnd_flows_dpillar_18_4.dat};

    % \addplot[only marks,\ficonncol,mark=square] table
    % {dat/rnd_flows_ficonn_2_24.dat};

\end{axis}
\end{tikzpicture}

%%% Local Variables: 
%%% mode: latex
%%% TeX-master: "GenDualPort"
%%% End: 
 \caption{Distribution of number of flows per link for the random traffic pattern.  Not plotted are  52,488 unused links in DPillar and 6,162 unused links in FiConn.
  %GQ$^\star_{3,10}$, GQ$^\star_{4,6}$, and FiConn$_{2,24}$,
  %DPillar$_{18,4}$ SP, respectively.
	}
\label{plot:rnd_flows}
\end{figure}
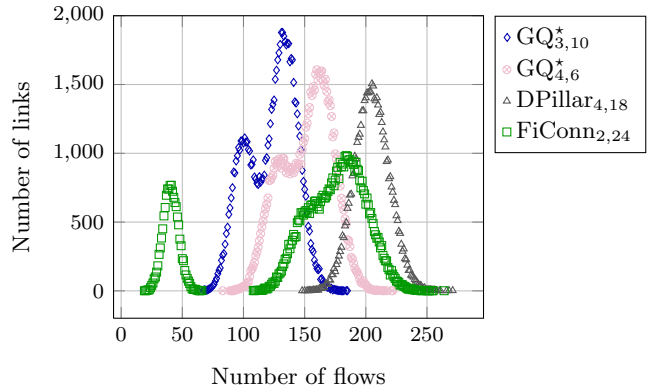

There are a number of open questions arising from this paper that we
will investigate in the future.  A non-comprehensive list is as
follows: analyse the practicalities (floor planning, wiring,
availability of local routing, and so on) of packaging the DCNs GQ$^\star$; perform a
broader evaluation using a higher number of DCN architectures and
traffic models; refine \emph{GQ$^\star$-routing\/} to produce minimal paths for
fault-free networks and compare its performance with the near-optimal
algorithm used in this paper; apply the stellar transformation to
other well-understood interconnection networks (some of which we have already highlighted); and, finally, explore the effect of
the stellar construction on formal notions of symmetry in the base
graph and in relation to metrics such as bisection width.

\section*{Acknowledgements }
This work has been funded by the Engineering and Physical Sciences
Research Council (EPSRC) through grants EP/K015680/1 and EP/K015699/1. Dr. Javier Navaridas is also supported by the European Union's Horizon 2020 programme under grant agreement No. 671553 `ExaNeSt'. The authors gratefully acknowledge their support.
 
\FloatBarrier
%\section*{References}
\bibliographystyle{abbrv}
%\bibliographystyle{elsarticle-num}
% plain}

%

 %

% \appendix
% \renewcommand{\thesection}{A}
% \setcounter{figure}{0} \renewcommand{\thefigure}{A.\arabic{figure}}
% \setcounter{table}{0} \renewcommand{\thetable}{A.\arabic{table}}

\end{document}